\documentclass{iopjournal}
\usepackage[T1]{fontenc}
\usepackage{lmodern}
\usepackage{color}

\newcommand \be{\begin{eqnarray}}
\newcommand \ee{\end{eqnarray}}
\usepackage{amsmath,amsfonts}
\usepackage{mathtools}
\usepackage{multirow}
\usepackage{makecell}
\usepackage{comment}
\usepackage{ragged2e}
\usepackage[normalem]{ulem}
\usepackage{cite}
\usepackage{caption}
\usepackage{cellspace}
\DeclareMathOperator{\Tr}{Tr}

\DeclareMathOperator{\diag}{diag}

\newcommand{\bea}{\begin{eqnarray}}
\newcommand{\eea}{\end{eqnarray}}
\newcommand{\beq}{\begin{equation}}
\newcommand{\eeq}{\end{equation}}
\newcommand{\bal}{\begin{equation}\begin{aligned}}
\newcommand{\eal}{\end{aligned} \end{equation}}

\newcommand{\cD}{{\mathcal D}}

\newcommand{\cL}{{\mathcal L}}

\newcommand{\cN}{{\mathcal N}}
\newcommand{\cP}{{\mathcal P}}

\newcommand{\cT}{{\mathcal T}}

\newcommand{\cO}{{\mathcal O}}

\newmuskip\pFqmuskip

\newcommand*\pFq[6][8]{%
  \begingroup % only local assignments
  \pFqmuskip=#1mu\relax
  \mathchardef\normalcomma=\mathcode`,
  \mathcode`\,=\string"8000
  \begingroup\lccode`\~=`\,
  \lowercase{\endgroup\let~}\pFqcomma
  {}_{#2}F_{#3}{\left[\genfrac..{0pt}{}{#4}{#5};#6\right]}%
  \endgroup
}
\newcommand{\pFqcomma}{{\normalcomma}\mskip\pFqmuskip}

\begin{document}
\justifying

\title{Strong coupling dynamics of defect RG flows in ABJM
%Defect RG flows in ABJM at strong coupling 
}

\author{Marco S. Bianchi$^1$, Luigi Castiglioni$^2$, Silvia Penati$^2$\begingroup
\renewcommand{\thefootnote}{\fnsymbol{footnote}}
\footnote[1]{Corresponding author. Email: silvia.penati@mib.infn.it.}
\endgroup, Marcia Tenser$^3$, Diego Trancanelli$^4$ }

\affil{$^1$Facultad de Ingeniería, Universidad San Sebastián, Santiago, Chile}
\vskip 5pt

\affil{$^2$Dipartimento di Fisica, Universit\`a degli Studi di Milano--Bicocca and INFN, Sezione di Milano--Bicocca,\\ Piazza della Scienza 3, 20126 Milano, Italy}
\vskip 5pt

\affil{$^3$International Institute of Physics, Federal University of Rio Grande do Norte, Campus Universit\'ario,\\Lagoa Nova, Natal-RN 59078-970, Brazil}
\vskip 5pt

\affil{$^4$Dipartimento di Scienze Fisiche, Informatiche e Matematiche, Universit\`a di Modena e Reggio Emilia, \\  via G. Campi 213/A, 41125 Modena, Italy and INFN Sezione di Bologna, via Irnerio 46, 40126 Bologna, Italy}

\email{marco.bianchi@uss.cl, l.castiglioni8@campus.unimib.it, silvia.penati@mib.infn.it, \\ marciatenser@gmail.com, diego.trancanelli@unimore.it} 

\begin{abstract}
\noindent
Wilson loop operators in ABJM theory provide a rich arena for studying defect conformal field theories (dCFTs) and the renormalization group (RG) flows connecting them. While these are well understood at weak coupling, a complete strong-coupling picture remains an open problem. 
In this paper, we present a systematic strong-coupling analysis of the fixed points and operator spectrum underlying defect RG flows in ABJM, via holography.
By examining fluctuations of fundamental strings in the AdS$_4 \times \mathbb{CP}^3$ background around classical AdS$_2$ solutions, we map worldsheet excitations to the operators in the dual dCFT which are responsible for the flows and determine their scaling dimensions, including subleading corrections from one-loop worldsheet effects. We show how different boundary conditions on string coordinates correspond to distinct operators and provide a geometric realization of the RG flows through interpolating boundary conditions. We apply this framework to fermionic 1/2 BPS, bosonic 1/6 BPS, and non-supersymmetric Wilson loops, establishing a coherent strong-coupling picture in which the 1/2 BPS loop is IR stable, the 1/6 BPS loop acts as a saddle point, and the non-supersymmetric configuration emerges as a natural UV fixed point. We also advance a proposal for the holographic dual of a second non-supersymmetric loop, in terms of averaging over Dirichlet boundary conditions.
\end{abstract}

\tableofcontents

\vskip 20pt

\section{Introduction}
\vskip 7pt

Wilson loop operators play a central role in the study of gauge theories, providing non-local probes that capture both perturbative and non-perturbative aspects of quantum field dynamics. In supersymmetric theories, they offer a particularly rich arena where exact results \cite{erickson,gross}, localization techniques \cite{pestun,Kapustin:2009kz,Marino:2009jd,Drukker:2010nc}, integrability \cite{Correa:2012at,Correa:2012hh}, and holography \cite{Maldacena_1998,Rey:1998ik} can be brought together. 

In three-dimensional $\mathcal{N}=6$ Chern-Simons-matter theories of ABJM type \cite{Aharony:2008ug,Aharony:2008gk}, Wilson loops exhibit a wide variety of realizations \cite{Drukker:2019bev}, ranging from purely bosonic operators to configurations preserving different fractions of supersymmetry, obtained by coupling the gauge connection to scalar and fermionic matter fields \cite{Berenstein:2008dc,Drukker:2008zx,Chen:2008bp,Rey:2008bh,Drukker:2009hy}. A remarkable feature of the existence of multiple Wilson loop operators is that they can be interpreted as different conformal defects supporting one-dimensional conformal field theories (CFT$_1$) \cite{Drukker:2006xg,Sakaguchi:2007ba,Cooke:2017qgm}. These defect theories are connected by renormalization group (RG) flows \cite{Polchinski:2011im}, triggered by marginally relevant deformations of the UV fixed point and driving the flow to IR conformal defects. At weak coupling, the structure of these flows has been elucidated in detail \cite{Castiglioni:2022yes,Castiglioni:2023uus}, revealing a network of fixed points corresponding to distinct Wilson loop operators, including the $1/2$ BPS, $1/6$ BPS, and non-supersymmetric configurations.

Understanding this structure at strong coupling is a natural and important step in completing the picture. Through the AdS/CFT correspondence, Wilson loops in ABJM theory admit a dual description in terms of fundamental strings propagating in the ${\rm AdS}_4 \times \mathbb{CP}^3$ background. In this framework, different loop operators are distinguished by the boundary conditions imposed on the string endpoints in the internal $\mathbb{CP}^3$ space. Deformations of the defect theory correspond to modifications of these boundary conditions, providing a geometric realization of RG flows in terms of interpolating string configurations \cite{Polchinski:2011im}.

In this paper, we develop a systematic strong-coupling analysis of the fixed points and operator spectrum underlying defect RG flows in ABJM theory.
Our strategy is to identify the defect operators responsible for these flows by studying fluctuations of the dual string worldsheet around classical ${\rm AdS}_2$ solutions. Exploiting the AdS$_2$/CFT$_1$ correspondence, we map worldsheet excitations to operators in the defect theory and determine their scaling dimensions \cite{Giombi:2017cqn}.
In principle, a complete treatment would require the full Green-Schwarz superstring in AdS$_4\times\mathbb{CP}^3$. However, the quantities explicitly computed in this work can be determined without a direct analysis of the fermionic sector. In the protected cases considered here, supersymmetry constrains the full result in such a way that fermionic contributions are not needed, and they can be inferred indirectly from multiplet structure and protection arguments. For unprotected operators, on the other hand, fermionic fluctuations may contribute non-trivially to anomalous dimensions and operator mixing. Nevertheless, all quantities computed explicitly in this paper are obtained at an order where such corrections do not contribute. A complete determination of the spectrum and mixing of unprotected composite operators would instead require a genuine one-loop analysis of the full Green-Schwarz theory, which lies beyond the scope of the present work.

A key aspect of our analysis is the interplay between boundary conditions and operator dimensions. Dirichlet and Neumann boundary conditions on the string coordinates correspond to different quantizations of the dual fields, leading to distinct spectra of defect operators. More generally, mixed (interpolating) boundary conditions provide a holographic realization of RG flows, connecting different fixed points along the radial direction of AdS. This perspective allows us to translate the field-theoretic picture of defect deformations into a geometric framework.

Although in most cases the classical scaling dimension already suffices to determine the RG behavior of the defect deformations, we go beyond this leading analysis and, whenever possible, compute the subleading corrections by extracting one-loop anomalous dimensions via worldsheet perturbation theory \cite{Giombi:2017cqn,Beccaria:2017rbe,Beccaria:2019dws}.

We apply this approach to a set of four Wilson loop operators: the fermionic $1/2$ BPS loop $W_{1/2}^+$, the bosonic $1/6$ BPS loop $W_{1/6}$, and the non-supersymmetric loops $W^\pm$. For each case, we identify the corresponding worldsheet fluctuations, construct suitable composite operators, and compute their scaling dimensions at strong coupling. This enables us to determine whether the associated deformations are relevant or irrelevant, and compare the resulting RG behavior with the expectations from weak-coupling analyses.

Our results provide a coherent holographic picture of defect RG flows in ABJM theory. In particular, we find that the $1/2$ BPS loop is an IR-stable fixed point, with all singlet deformations corresponding to irrelevant operators at strong coupling, while the $1/6$ BPS loop exhibits both relevant and irrelevant directions, consistent with its role as a saddle point in the space of flows. The non-supersymmetric loop $W^-$ emerges as a natural UV fixed point, from which flows are triggered by relevant scalar deformations. 

More generally, our analysis highlights how the structure of defect RG flows is encoded in the spectrum of string fluctuations and their boundary conditions, providing a direct link between field-theoretic deformations and geometric data in the dual string theory. This work extends previous studies of defect CFTs and Wilson loops in higher-dimensional theories~\cite{Polchinski:2011im,Giombi:2017cqn,Beccaria:2017rbe,Beccaria:2019dws}, and opens the way to further investigations of non-supersymmetric defects, operator mixing, and the role of fermionic modes in the strong-coupling regime.

The paper is organized as follows. In section \ref{sec:general} we review notable Wilson loop operators in ABJM theory, their nature under RG flows as emerged from previous weak-coupling analyses, and their string dual descriptions, fixing our conventions and discussing the relevant classical solutions. In section \ref{sec:W12+} we analyze the spectrum of fluctuations around the AdS$_2$ solution corresponding to the $1/2$ BPS Wilson loop, with particular emphasis on their boundary conditions and their interpretation in terms of defect operators. We establish a dictionary between string modes and operators in the defect CFT, identifying the relevant composite operators through their symmetry properties and determining their dimensions up to subleading order via worldsheet perturbation theory. In section \ref{sec:W16} we focus on the $1/6$ BPS Wilson loop and study its deformations at strong coupling, computing anomalous dimensions and comparing with the corresponding weak-coupling picture of the RG flows. In section \ref{sec:Wminus} we analyze the $W^{-}$ defect and its role as a UV fixed point, discussing the structure of deformations and the boundary conditions in the string description. In section \ref{sec:Wplus} we introduce a novel holographic description of the IR fixed point given by the other non-BPS Wilson loop, $W^+$. Finally, in section \ref{sec:conclusions} we summarize our results and outline possible future directions.
\section{General features of Wilson loops in ABJM and their string duals}
\label{sec:general}
\vskip 7pt

In three dimensions, ${\cal N} \geq 2$ supersymmetric Chern-Simons-matter theories \cite{Gaiotto:2008sd, Imamura:2008dt, Hosomichi:2008jd, Hama:2010av} allow for the construction of a plethora of Wilson-like operators (see, for example,  \cite{Gaiotto:2007qi,Drukker:2008zx,Drukker:2009hy,Cardinali:2012ru, Ouyang:2015qma,Cooke:2015ila,Ouyang:2015bmy, Mauri:2017whf, Mauri:2018fsf, Drukker:2020dvr, Drukker:2022ywj, Drukker:2022bff, Drukker:2022txy, Castiglioni:2022yes} and \cite{Drukker:2019bev} for a review), thanks to the possibility of completing the ordinary gauge holonomy with scalar bilinears and fermions. Their general structure has the form 
\begin{equation}
    W = \Tr\cP\exp\left(-i\int_\mathcal{C} \cL\, d\tau \right),
\end{equation}
where $\mathcal{C}$ is the curve that supports the loop and $\tau$ its affine parameter. In what follows we will take the trace in the fundamental representation of the gauge group and $\mathcal{C}$ to be a circle, with $\tau$ being its angular coordinate
\begin{equation}
\mathcal{C}: \quad x^\mu=(0,\cos\tau,\sin\tau)\,.  
\end{equation}

Specifying this construction to the ABJM theory \cite{Aharony:2008ug}, the quantity $\cL$ is a $U(N|N)$ supermatrix\footnote{Here we use notations and conventions of \cite{Castiglioni:2023uus}.},
\begin{equation}\label{eq:cL}
    \cL = \begin{pmatrix}
    A_\mu \dot{x}^\mu -\frac{2\pi i}{k}\vert\dot{x}\vert M_I^J C_J\bar{C}^I && -i\sqrt{\frac{2\pi}{k}}\vert\dot{x}\vert \eta_I^\alpha\bar\psi^I_\alpha \\
    -i\sqrt{\frac{2\pi}{k}}\vert\dot{x}\vert \psi_I^\alpha\bar\eta_\alpha^I && \hat{A}_\mu \dot{x}^\mu - \frac{2\pi i}{k}\vert\dot{x}\vert M_I^J \bar{C}^I C_J
\end{pmatrix},
\end{equation}
where $M_I^J$ controls the coupling to scalar fields and $\eta_I^\alpha,\bar{\eta}^I_\alpha$ the coupling to fermions \cite{Drukker:2009hy}. In fact, distinct choices for these matter couplings give rise to different loops, which may preserve a fraction of the supercharges (BPS Wilson loops) \cite{Drukker:2008zx, Drukker:2009hy, Castiglioni:2022yes} or break supersymmetry completely \cite{Castiglioni:2023uus}. 

In this paper we focus on four operators, $W_{1/2}^+$, $W_{1/6}$, and $W^\pm$.  In table \ref{tab:WLs} we relate them to the corresponding matter couplings. Note that fermions do not appear in $W_{1/6}$ and  $W^\pm$, which are therefore called bosonic loops. For these operators it is sufficient to consider only the contribution from the first node of the ABJM quiver, {\em i.e.} the upper diagonal term in \eqref{eq:cL}, since the expectation value of the lower-right contribution can be obtained by complex conjugation.
\begin{table}[ht]
\centering
\begin{tabular}{|c|c|c|}
\hline
\multirow{3}{*}{$W_{1/2}^+$}
& \multirow{3}{*}{$M_I^J=\diag(-1,1,1,1)$}
& \multirow{3}{*}{\makecell{$\eta_I^\alpha=\begin{pmatrix}
    e^{i\tau/2} & -ie^{-i\tau/2}
\end{pmatrix}\delta_I^1\,,\quad \bar{\eta}^I_\alpha=\begin{pmatrix}
    ie^{-i\tau/2} \\ -e^{i\tau/2}
\end{pmatrix}\delta^I_1$}}  \\
& & 
\\
& &  \\
\hline
\multirow{2}{*}{$W_{1/6}$} & \multirow{2}{*}{$M_I^J=\diag(-1,-1,1,1)$} & \multirow{2}{*}{$\eta_I^\alpha=\bar{\eta}^I_\alpha=0$}  \\
& & 
\\
\hline
\multirow{2}{*}{$W^\pm$} & \multirow{2}{*}{$M_I^J=\pm\mathbf{1}$} & \multirow{2}{*}{$\eta_I^\alpha=\bar{\eta}^I_\alpha=0$} \\
& & 
\\
\hline
\end{tabular}
\caption{\label{tab:WLs} Wilson loops considered in this paper and their corresponding matter couplings.}
\end{table}

The $W_{1/2}^+$ and $W_{1/6}$ operators are 1/2 and 1/6 BPS, and preserve $SU(3)$ and $SU(2) \times SU(2)$ subgroups of the $SU(4)$ R-symmetry, respectively. Their string theory duals at strong coupling have been determined before, see for instance \cite{Chen:2008bp,Drukker:2008zx,Rey:2008bh,Drukker:2009hy} and the review in sections 12-13 of  \cite{Drukker:2019bev}. We do, however, shed new light on their dual descriptions. The other two operators, $W^\pm$, are not supersymmetric and preserve the entire $SU(4)$ R-symmetry group. They were first discussed in \cite{Castiglioni:2023uus} in the context of defect RG flows. Their strong coupling description is completely new.

%%%%%%%%%%%%%%%%%%

\subsection{The RG flows}
\vskip 7pt

\begin{figure}[ht]
    \centering \includegraphics[width=.4\textwidth]{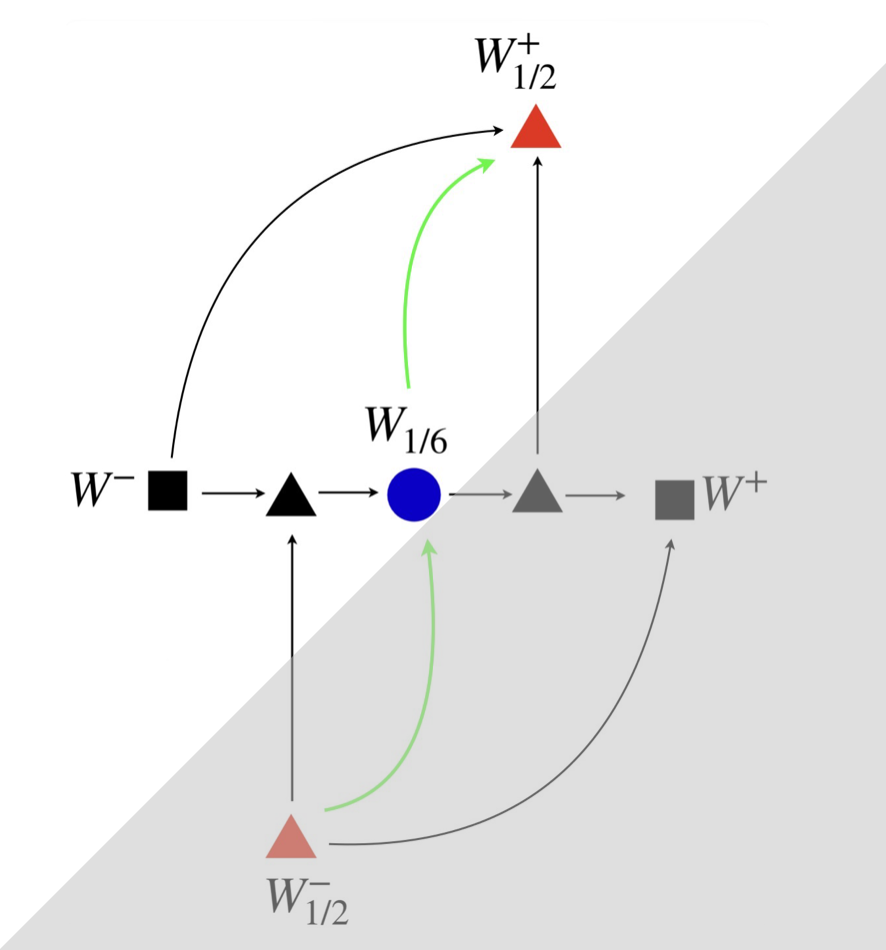}
    \caption{Representation of the RG flows connecting different Wilson loop operators in ABJM theory. Arrows point towards the IR. Along the horizontal line we have purely bosonic deformations whereas in the vertical direction we turn on fermions. Mixed deformations correspond to compositions of these two. Black triangles represent $SU(3)$ bosonic operators that will not be included in the present discussion. Green lines represent enriched flows. The gray shaded area covers what we call ``mirror Wilson loops''. We only discuss $W^+$ in some detail in section \ref{sec:Wplus}, while we comment on them in the discussion section.
    }
\label{fig:fluxes}
\end{figure}

As largely discussed in \cite{Castiglioni:2022yes, Castiglioni:2023uus} and reviewed in \cite{Penati:2025mrf}, the four Wilson loop operators under consideration sit at the fixed points of RG flow trajectories, driven by marginally relevant deformations. The network of RG flows is schematically depicted in figure \ref{fig:fluxes}, where arrows point towards the IR. Horizontal lines represent bosonic flows, while vertical lines and curves include fermions. 

In short, the figure indicates that $W^-$ and $W^-_{1/2}$ are UV unstable fixed points, whereas under perturbations preserving some fraction of the R-symmetry group the system is naturally driven towards  $W^+_{1/2}$ and $W^+$, which are then IR stable fixed points. The bosonic $W_{1/6}$ operator is instead a saddle point. More precisely, $W^-$ is UV unstable under both  bosonic and mixed bosonic-fermionic marginally relevant deformations that preserve $SU(3)$ R-symmetry. While the mixed deformation drives the system towards the $W^+_{1/2}$ IR stable fixed point, along the bosonic direction it can be further perturbed by an $SU(3) \to SU(2) \times SU(2)$ breaking deformation to reach $W_{1/6}$, which in turn can be perturbed by an $SU(4)$ restoring deformation to reach $W^+$. Similarly, $W^-_{1/2}$ represents a UV unstable fixed point from which we can flow to $W_{1/6}, W^+$ or another bosonic $SU(3)$ preserving operator (the left black triangle in figure \ref{fig:fluxes}), depending on which $SU(4)$ subgroup is preserved by the perturbation. In general, no supersymmetry is preserved along the flows, except for the special $W^-_{1/2} \to W_{1/6} \to W^+_{1/2}$ flows drawn in green, which are induced by marginally relevant perturbations preserving 1/6 of the supercharges. For this reason, these special RG trajectories were dubbed ``enriched flows'' in \cite{Castiglioni:2022yes}.

%%%%%%%%%%%%%%%%%%%

\subsection{The dual string}
\vskip 7pt

The dual description of ABJM theory is given by M-theory in AdS$_4\times S^7/\mathbb{Z}_k$, where $k$ is the Chern-Simons level. Writing $S^7$ as a circle fibration over $\mathbb{CP}^3$, for large enough $k$ the radius of the M-theory circle shrinks and the theory is effectively described as a type IIA string theory on AdS$_4 \times \mathbb{CP}^3$. The ten-dimensional metric is
\begin{equation}
     ds^2 = ds^2_{{\rm AdS}_4} + 4 ds^2_{\mathbb{CP}^3}\,,
\end{equation}
where we have set the AdS radius to one. The other background fields are 
\begin{equation}
    e^{2\phi} = \frac{4}{k^2}\,, \qquad F^{(4)} =\frac{3}{2}k \, \text{vol}(\text{AdS}_4) \,,\qquad F^{(2)} = \frac{k}{4} dA,
\end{equation}
where $\phi$ is the dilaton, $\text{vol}(\text{AdS}_4)$ is the AdS volume and $dA$ is the K\"ahler form on $\mathbb{CP}^3$. 

Different Wilson loop operators in the fundamental representation correspond to fundamental strings ending on the loop contour at the AdS boundary and distinguished by their boundary conditions in the internal $\mathbb{CP}^3$ space. In order to describe the dual string solutions, it is convenient to use a Poincaré patch of the metric for the AdS$_4$ factor
\begin{equation}\label{eq:ads4}
    ds^2_{{\rm AdS}_4} = \frac{dz^2 + dx^A dx^A}{z^2},
\end{equation}
where $z$ is the radial coordinate and $x^A=(x^0,x^1,x^2)$ parameterize the Euclidean three-dimensional boundary located at $z=0$. 

For the $\mathbb{CP}^3$ factor, we can use either homogeneous coordinates or the affine parametrization, according to the R-symmetry structure that we want to be explicitly realized, which in turn is dictated by the amount of R-symmetry preserved by the dual operator that we want to describe. 

On the one hand, by writing $\mathbb{CP}^3$ as a subspace of the 7-sphere written in complex coordinates, we introduce homogeneous coordinates $Z_I$ ($I=1,2,3,4$) satisfying $Z_I = e^{i\phi}Z_I$, for any real parameter $\phi$, and $Z_I\bar Z^I=1$. The metric reads
\begin{equation}
ds^2_{\mathbb{CP}^3} = \frac{dZ_I d\bar Z^I}{Z_J \bar Z^J} - \frac{Z_I\bar Z^Jd Z^Id\bar Z_J}{|Z_K \bar Z^K|^2} \, .
\end{equation}
Such a set of coordinates makes the $SU(4)$ structure of $\mathbb{CP}^3$ explicit.

On the other hand, if we are interested in the dual description of a Wilson loop that preserves only the $SU(3) \subset SU(4)$ subgroup, it is more convenient to work with the affine parametrization. We pick one of the four homogeneous coordinates, say $Z_1$, define $\omega_i= Z_i/Z_1$ ($i=2,3,4$) and remove the $S^1$ fiber by imposing the constraint $\omega_i\bar\omega^i +1 = 1/Z_1\bar Z^1$. The metric in these coordinates reads
\begin{equation}\label{eq:affine}
    ds^2_{\mathbb{CP}^3} = \frac{d\omega_id\bar\omega^i}{1+\omega_j\bar\omega^j} - \frac{\omega_i\bar\omega^jd\bar\omega^id\omega_j}{(1+\omega_k\bar\omega^k)^2} \, .
\end{equation}

Finally, if we are interested in solutions that preserve some $SU(2)$ internal symmetry, it is more convenient to parametrize the homogeneous coordinates in terms of Hopf ones, by setting
\begin{equation}
\label{eq:hopf}
\begin{aligned}
Z_1 &= \cos\frac{\alpha}{2}\cos\frac{\theta_1}{2}
      e^{\frac{i}{2}\varphi_1}e^{\frac{i}{4}\chi},
&\qquad
Z_2 &= \cos\frac{\alpha}{2}\sin\frac{\theta_1}{2}
      e^{-\frac{i}{2}\varphi_1}e^{\frac{i}{4}\chi},\\
Z_3 &= \sin\frac{\alpha}{2}\cos\frac{\theta_2}{2}
      e^{\frac{i}{2}\varphi_2}e^{-\frac{i}{4}\chi},
&
Z_4 &= \sin\frac{\alpha}{2}\sin\frac{\theta_2}{2}
      e^{-\frac{i}{2}\varphi_2}e^{-\frac{i}{4}\chi}.
\end{aligned}
\end{equation}
with $0\leq\alpha,\theta_1,\theta_2\leq\pi,\,
0\leq\varphi_1,\varphi_2<2\pi,\,0\leq\chi<4\pi$. The metric in angular coordinates then reads 
\begin{equation}\label{eq:cp3angular}
\begin{split}
ds^2_{\mathbb{CP}^3}=\frac14\bigg[&d\alpha^2
+\cos^2\frac{\alpha}{2}\left(d\theta_1^2+\sin^2\theta_1\,d\varphi_1^2\right)
+\sin^2\frac{\alpha}{2}\left(d\theta_2^2+\sin^2\theta_2\,d\varphi_2^2\right)\\
&+\sin^2\frac{\alpha}{2}\cos^2\frac{\alpha}{2}
\left(d\chi+\cos\theta_1\,d\varphi_1-\cos\theta_2\,d\varphi_2\right)^2\bigg].
\end{split}
\end{equation}

\vskip 5pt
The string solutions dual to Wilson loop operators always satisfy Dirichlet boundary conditions in AdS$_4$, typically realized by choosing the static gauge $x^0 = \tau$ and $z = \sigma$ where $(\tau,\sigma)$ are the string worldsheet coordinates, and setting $x^1|_{z=0} = x^2|_{z=0} = 0$. The minimal area surface satisfying these boundary conditions corresponds to $x^1(\tau,\sigma) =x^2(\tau,\sigma)=0$ and describes an AdS$_2$ inside AdS$_4$ that ends at the boundary on the Wilson loop contour parametrized by $\tau$. It is convenient to embed the AdS$_2$ solution inside 
AdS$_4$ as 
\begin{equation}
\label{eq:AdS2solution}
ds^2_{\rm {AdS_4}} = \frac{(1+\frac{x^2}{4})^2}{(1-\frac{x^2}{4})^2} ds^2_{\rm {AdS}_2} + \frac{d x^a dx^a}{(1-\frac{x^2}{4})^2}\, ,
\qquad \quad ds^2_{\rm {AdS}_2} = \frac{1}{\sigma^2} (d\sigma^2 + d\tau^2)\,,
\end{equation}
where $a=1,2$ labels the coordinates transverse to the AdS$_2$ solution and $x^2 \equiv x^a x_a$.

This solution is common to all Wilson loops. What discriminates between different operators  preserving different amounts of supersymmetry is instead the set of boundary conditions on the internal $\mathbb{CP}^3$ coordinates. As we are going to review and further elaborate later, maximal supersymmetry, that is the 1/2 BPS operator, is obtained by setting Dirichlet boundary conditions in all the six internal directions \cite{Drukker:2009hy}, whereas trading some Dirichlet boundary conditions for Neumann ones and smearing along these directions reduces the amount of supersymmetry \cite{Drukker:2008zx}. As we are going to show in section \ref{sec:Wplus}, one exception is represented by the non-supersymmetric $W^+$ operator, whose dual description can be obtained by smearing over Dirichlet boundary conditions. 

In this framework, interpolating boundary conditions provide a holographic realization of the RG flows connecting the corresponding fixed points \cite{Witten:2001ua,Polchinski:2011im}
\begin{equation}
\label{eq:bc}
    Z_I(\tau, \sigma=0) + f(z) \, n^a \partial_a Z_I(\tau, \sigma=0) =0\,,
\end{equation}
where $n^a$ is a unit vector orthogonal to the boundary and the interpolating parameter $f$ depends on the radial AdS coordinate in such a way that $f(z=0) = \infty$ (UV fixed point) and $f(z \to \infty) = 0$ (IR fixed point). String solutions subject to these mixed boundary conditions are dual to families of Wilson loops obtained by marginally relevant deformations of operators at the UV fixed points. Particular cases of interpolating boundary conditions that preserve supersymmetry or scale symmetry have been studied in \cite{Correa:2019rdk,Garay:2022szq,Correa:2023thy}. In this paper we are interested in more general interpolating conditions of the form \eqref{eq:bc}, where the dimensionful parameter $f$ breaks (super)conformal invariance, thus providing the dual of RG flows in the CFT. A similar investigation has been carried out for ${\cal N}=4$ super Yang-Mills theory in \cite{Beccaria:2017rbe}.

One could in principle generalize the solution to the ABJ case, where the string couples to additional background fluxes, in particular the $B$-field. Such couplings are expected to account for framing effects at strong coupling \cite{Bianchi:2024sod,Bianchi:2025xry}. In this work, however, we restrict to the purely geometric embedding and do not include these contributions.

In the rest of the paper we analyze the string dual of the one-dimensional CFTs defined by the Wilson loops in table \ref{tab:WLs}, with the aim of identifying the holographic duals of defect operators that trigger the RG flows away from the fixed points. Our strategy is to match boundary operators in the one-dimensional defect CFT with fluctuations of the dual open string in $\text{AdS}_4 \times \mathbb{CP}^3$. In particular, we focus on operators that are singlets under the preserved R-symmetry, since these are the natural candidates to generate irrelevant deformations. By expanding the Wilson loop couplings in terms of small fluctuations of the $\mathbb{CP}^3$ coordinates, we will identify the corresponding composite worldsheet fields and compute their anomalous dimensions at strong coupling.
\section{The fermionic \texorpdfstring{$1/2$ BPS loop $W_{1/2}^+$}{1/2 BPS loop W1/2+}}
\label{sec:W12+}
\vskip 7pt

We begin by considering the most supersymmetric defect in ABJM theory, that is the $1/2$ BPS Wilson loop $W_{1/2}^+$ defined as in \eqref{eq:cL} with the couplings given in the first line of table \ref{tab:WLs}. We consider deformations of this defect by classically marginal, gauge invariant operators. In the superconnection construction, these can be naturally constructed as scalar bilinears on the diagonal entries 
\begin{equation}
    \frac{4\pi i}{k} \begin{pmatrix}
        C_I\bar C^J & 0 \\ 0 &  \bar C^J C_I
    \end{pmatrix}, \qquad \qquad I, J = 1,2,3,4,
\end{equation}
or as fermionic contributions in the off-diagonal ones\footnote{The spinor indices, which we suppress, are always contracted as in \eqref{eq:cL}.}
\begin{equation}
    i\sqrt{\frac{2\pi}{k}}
    \begin{pmatrix}
        0 & \eta \bar\psi^I \\  \psi_I \bar\eta & 0
    \end{pmatrix} , \qquad \qquad I= 1,2,3,4\,.
\end{equation}
The defect preserves an $SU(3)$ R-symmetry subgroup of the ABJM global symmetry, which allows defect operators to be organized into irreducible representations of $SU(3)$. In particular, $SU(3)$ singlet operators read (the sign choices are for later convenience)
\begin{equation}
\label{eq:singlets}
    S_1 = -\frac{4\pi i}{k} \begin{pmatrix}
         C_1\bar C^1 & 0 \\ 0 &  \bar C^1 C_1
    \end{pmatrix},\qquad S_2 = \frac{4\pi i}{k}
    \begin{pmatrix}
         C_i\bar C^i & 0 \\ 0 &  \bar C^i C_i
    \end{pmatrix}, \qquad  S_3 =i\sqrt{\frac{2\pi}{k}}  \begin{pmatrix}
        0 & \eta \bar\psi^1 \\  \psi_1 \bar\eta & 0
    \end{pmatrix} ,
\end{equation}
where, according to our definitions, we selected the preferred direction 1 for symmetry breaking. The operators transforming in the fundamental representation of $SU(3)$ can be combined to form the operator
\begin{equation}\label{eq:Oi}
    \mathbb{O}^i = \begin{pmatrix} \frac{4\pi i}{k}  C_1 \bar C^i & -i \sqrt{\frac{2\pi}{k} } \eta \bar\psi^i \\ 0 & \frac{4\pi i}{k} \bar C^i C_1  \end{pmatrix}, \qquad \qquad i=2,3,4,
\end{equation}
and its hermitian conjugate, which belong to the displacement multiplet and have protected dimension one. We will ignore unprotected operators in the fundamental, as they are not relevant for the present discussion. The remaining scalar bilinears
\begin{equation}
\label{eq:traceless}
    T_i{}^j  =  \frac{4\pi i}{k} \begin{pmatrix}
        C_i\bar C^j & 0 \\ 0 &  \bar C^j C_i
    \end{pmatrix}, \qquad \qquad i \ne j = 2,3,4,
\end{equation}
transform in the adjoint representation of $SU(3)$.

The unprotected operators considered so far possess anomalous dimensions that can potentially trigger an RG flow for which $W_{1/2}^+$ acts as a fixed point. Indeed, the singlet operators in \eqref{eq:singlets} were analyzed at weak coupling in \cite{Castiglioni:2023uus} and were found to correspond to marginally irrelevant deformations, with $W_{1/2}^+$ emerging as a stable IR attractor. 

Precisely, let us consider deformations $\chi_1 S_1$, $\chi_2 S_2$ and $\chi_3 S_3$ around $W_{1/2}^+$.
They correspond to perturbations of the superconnection appearing in the path-ordered exponential,
\begin{equation}
W[\chi_i]
=
\frac{1}{N}\,
{\rm Tr}\,
P\exp\!\left(
-i\int d\tau\,
\Big(\mathcal L_{W^+_{1/2}}
+\sum_{i=1}^{3}\chi_i\,S_i
\Big)
\right).
\end{equation}
The family of Wilson loops is parameterized so that $\chi_i=0$ corresponds to the IR fixed point $W_{1/2}^+$, while $\chi_i=1$ corresponds to the UV fixed point $W^-$. Thus, in this shifted parametrization the actual infinitesimal couplings around the IR fixed point are the $\chi_i$ themselves.

The three parameters undergo a non-trivial renormalization, which at one loop in the coupling constant leads to the following $\beta$-functions 
\begin{equation}\label{eq:beta}
\begin{split}
    \beta_{\chi_1} &= \frac{N}{k} \,(\chi_1-1+(1-\chi_3)^2)\chi_1 + {\cal O} \left(\frac{N^2}{k^2} \right)\,,  \\
    \beta_{\chi_2} &= \frac{N}{k} \, (2+(\chi_3-2)\chi_3-\chi_2)\chi_2 + {\cal O} \left(\frac{N^2}{k^2} \right)\,, \\
    \beta_{\chi_3} &= \frac{N}{k} \, (1-\chi_3)(2-\chi_3)\chi_3 + {\cal O} \left(\frac{N^2}{k^2} \right)\,.
\end{split}
\end{equation}
These expressions were derived in \cite{Castiglioni:2023uus,Penati:2025mrf}, considering deformations of the UV fixed point $W^-$. Here we rewrote them as deformations of the IR fixed point $W_{1/2}^+$, shifting the couplings in such a way that this point corresponds to $\chi_i=0$. The corresponding scaling dimensions of the singlets read up to one loop
\begin{equation}
    \Delta_{S_1} = 1 + {\mathcal O}\left(\frac{N^2}{k^2}\right) \,, \qquad \Delta_{S_2} = 1 + 2\frac{N}{k} + {\mathcal O}\left(\frac{N^2}{k^2}\right) \,, \qquad \Delta_{S_3} = 1 + 2\frac{N}{k} + {\mathcal O}\left(\frac{N^2}{k^2}\right) \, .
\end{equation}
As anticipated, $S_2$ and $S_3$ correspond to marginally irrelevant operators at the IR fixed point associated with $W_{1/2}^+$. The absence of quantum corrections to the scaling dimension of $S_1$ is likely a one-loop artifact, which we expect to be lifted by higher-loop effects. It is therefore natural to expect $S_1$ to be marginally irrelevant as well.

Furthermore, although the operators do not mix at one-loop order, the $\beta$-functions for the corresponding deformation couplings already exhibit mixing of $\chi_3$ with $\chi_1$ and $\chi_2$. At higher orders we therefore expect all three singlet operators to mix non-trivially, with suitable linear combinations forming a basis of operators with definite scaling dimension.

A broader class of classically marginal deformations can also be considered, such as the scalar bilinears $T_i{}^j$ in \eqref{eq:traceless}, transforming in the adjoint representation of $SU(3)$. These operators generically correspond to scalar deformations associated with a non-diagonal scalar coupling matrix $M^I_{\;J} C_I \bar{C}^J$ in \eqref{eq:cL}. While we expect these deformations to be marginally irrelevant as well, it is less clear which UV fixed points the corresponding RG flows would originate from. Such fixed points might involve non-diagonal scalar coupling matrices, thus falling outside the general classification of fixed points characterized by constant scalar couplings \cite{Drukker:2022bff}. A more detailed investigation of these deformations would certainly be interesting, but we leave this question for future work.

We now turn to the study of $W_{1/2}^+$ at strong coupling. Our aim is to identify defect deformations that could be dual to (some of) the weak-coupling ones, determine whether they correspond to relevant or irrelevant perturbations, and relate them to previous results obtained in precision calculations \cite{Bianchi:2020hsz,Gorini:2022jws}.

At strong coupling we analyze defect deformations using the holographic dual description in terms of the string worldsheet sigma model. The $W_{1/2}^+$ defect is described by an open string whose worldsheet is an ${\rm AdS}_2$ surface embedded in ${\rm AdS}_4$, while being localized at a point in $\mathbb{CP}^3$ \cite{Drukker:2009hy}. This configuration preserves an $SO(2,1)\times SU(3)\times U(1)$ symmetry. Identifying the $SO(2,1)$ isometry group of ${\rm AdS}_2$ with the one-dimensional conformal group, and the $SU(3)$ subgroup of the $\mathbb{CP}^3$ isometries with the R-symmetry preserved by the defect, we recover precisely the symmetry group of $W_{1/2}^+$. We will therefore use the $SU(3)$ symmetry as a guiding principle to identify defect deformations at strong coupling that are dual to the field theory operators introduced above, in particular the $SU(3)$ singlets \eqref{eq:singlets}, whose RG properties are known at weak coupling.

Deformations of the defect theory are implemented as boundary terms added to the worldsheet action. These boundary interactions modify the variational problem and therefore the boundary conditions satisfied by the bulk fields, typically interpolating between Neumann and Dirichlet boundary conditions for the $\mathbb{CP}^3$ coordinates. To study the resulting operator spectrum and correlation functions, we expand the sigma model in fluctuations around the classical saddle.  Perturbative computations are then performed in terms of the fluctuation fields, whose propagators and interaction vertices follow from the expansion of the worldsheet action and the induced boundary conditions.

A complete treatment would require the full Green-Schwarz action in $AdS_4\times\mathbb{CP}^3$. For the quantities considered below, however, the bosonic fluctuation analysis is sufficient: protected contributions are fixed by supersymmetry, while the unprotected quantities we compute are insensitive to explicit fermionic corrections at the order considered.

We work with the affine parameterization of $\mathbb{CP}^3$, see eq. \eqref{eq:affine}, such that the Nambu-Goto action in the fixed static gauge reads \cite{Bianchi:2020hsz}
\begin{equation}\label{eq:Nambu-Goto1/2}
    S = T \int d\sigma d\tau \sqrt{\text{det}\left[ \frac{(1+\frac{x^2}{4})^2}{(1-\frac{x^2}{4})^2} g_{\mu\nu}+  \frac{\partial_\mu x^a \partial_\nu x^a}{(1-\frac{x^2}{4})^2}+ 4 \left( \frac{\partial_{\mu}\omega_i \partial_{\nu}\bar\omega^i}{1+|\omega|^2} - \frac{\omega_i\bar\omega^j \partial_{\mu}\omega_j \partial_{\nu}\bar\omega^i}{\left(1+|\omega|^2\right)^2} \right) \right]}\,,
\end{equation}
where $g_{\mu\nu}$ is the induced AdS$_2$ metric in \eqref{eq:AdS2solution} and the string tension is $T=\frac{\sqrt{\lambda}}{2\pi}$ (with $\lambda \equiv 2 \pi^2 N/k$), adopting the conventions of \cite{Correa:2023thy}. 

Expanding this action around the minimal area solution $x^1=x^2=0$, $\omega_i=0$ in powers of the transverse fluctuations $x^a$ ($a=1,2$) and $\omega_i$ ($i=2,3,4$), we obtain an interacting theory for two massive real scalar fields $x^a$ ($m^2=2$), associated with transverse fluctuations in ${\rm AdS}_4$, and three massless complex scalars $\omega_i$, describing fluctuations in $\mathbb{CP}^3$. The worldsheet action expanded up to quartic order can be found in \cite{Bianchi:2020hsz}.

According to the AdS$_2$/CFT$_1$ dictionary, scalar fields obey the relation $\Delta(\Delta-1)=m^2$, which relates the bulk mass to the scaling dimension of the dual defect operator. The scalar fluctuations in the ${\rm AdS}_4$ directions are therefore dual to operators with $\Delta=2$, while the $\omega_i$ and $\bar\omega^i$ fluctuations in $\mathbb{CP}^3$ correspond, in the standard quantization with Dirichlet boundary conditions \cite{Witten:2001ua}, to $\mathrm{CFT}_1$ operators of dimension $\Delta=1$. These operators have been identified in \cite{Bianchi:2020hsz} with bosonic components of the displacement operator supermultiplet on the Wilson loop, whose dimension is protected by supersymmetry. In particular, the $\mathbb{CP}^3$ fluctuations transforming in the fundamental representation of $SU(3)$ are dual to the operators $\mathbb{O}^i$ introduced in \eqref{eq:Oi}~\cite{Bianchi:2020hsz,Gorini:2022jws}. 

In the following we extend this approach to identify the string dual of non-protected operators, focusing in particular on $SU(3)$ singlets, which correspond to defect deformations and can trigger RG flows. An analogous investigation already exists for $\cN=4$ super Yang-Mills \cite{Giombi:2017cqn}, while for ABJM theory such an analysis is novel.  

In $\mathcal{N}=4$ super Yang-Mills the spectrum of defect operators associated with the $1/2$ BPS Wilson line can be analyzed at strong coupling using the dual string description in AdS$_5\times S^5$. The Wilson line is dual to a fundamental string whose worldsheet is an AdS$_2$ surface, and fluctuations of the string in the $S^5$ directions correspond to operator insertions on the $\mathrm{CFT}_1$ defect. Expanding the embedding coordinates of $S^5$ around the classical solution, one introduces five transverse fluctuation fields $y_a$, $a=1,\dots,5$, which transform in the fundamental representation of the preserved $SO(5)\subset SO(6)$ symmetry. These worldsheet fields are dual to elementary defect operators $\Phi_a$ with protected conformal dimension. In addition, the Wilson line itself contains an $SO(6)$ scalar coupling that selects a preferred direction on $S^5$, corresponding to the sixth component of the embedding coordinates. The associated defect operator $\Phi_6$ is an $SO(5)$ singlet and is not protected. In the string description it is identified at leading order with the bilinear $y_a y_a$, which is the simplest $SO(5)$ invariant combination that can be constructed from the worldsheet fields. Its dimension has been computed up to first order in worldsheet perturbation theory and corresponds to a dual irrelevant defect operator, in agreement with the 1/2 BPS Wilson loop being an IR fixed point.

Moving to the ABJM theory, the $\mathbb{CP}^3$ fluctuations $\omega_i$ play a similar role to $y_a$, corresponding, as recalled above, to protected operators in the $SU(3) \subset SU(4)$ fundamental representation.  We now identify unprotected operators, which are capable of probing the RG flow properties of the defect. 

The scalar couplings appearing in the loop operator determine precisely the internal orientation of the string endpoints in $\mathbb{CP}^3$. Writing them in terms of the homogeneous coordinates of $\mathbb{CP}^3$ \cite{Rey:2008bh,Correa:2014aga} makes the correspondence manifest: small variations of the couplings translate into fluctuations of the string coordinates and allow us to identify the defect operators they source. For $W_{1/2}^+$, we generalize this relation to the R-symmetry part of the fermionic couplings, writing\footnote{We split the fermionic couplings in table \ref{tab:WLs} as $\eta_I^\alpha = r_I \eta^\alpha$, with $\eta^\alpha \equiv (e^{i\tau/2}\,, -ie^{-i\tau/2})$ and $r_I \equiv \delta_I^1$.}  
\begin{equation}
    M_I^{J} = \delta_I^J - 2\frac{Z_I \bar Z^J}{Z_K \bar{Z}^K}\,, \qquad r_I = \frac{Z_I}{\sqrt{Z_K \bar{Z}^K}}\,.
\end{equation}
Switching to affine coordinates $Z_I = (1,\omega_i)$ for $i=2,3,4$, and expanding in small $\omega_i,\bar\omega^i$ fluctuations, we find
\begin{equation}\label{eq:M1}
    M_I^{J} = \delta_I^J -\frac{2}{1+\omega_k\bar\omega^k}\begin{pmatrix}
        1 & \bar\omega^j\\  \omega_i & \omega_i\bar\omega^j 
    \end{pmatrix} \simeq \delta_I^J -2\begin{pmatrix}
        1-\omega_k\bar\omega^k & \bar\omega^j\\  \omega_i & \omega_i\bar\omega^j 
    \end{pmatrix} + \cO(\omega^3)\,,
\end{equation}
and 
\begin{equation}\label{eq:f1}
    r_I = \begin{pmatrix}
        1 -\frac{1}{2}\omega_k\bar\omega^k \\ \omega_i
    \end{pmatrix} + \cO(\omega^3)\,.
\end{equation}

Inserting \eqref{eq:M1} and \eqref{eq:f1} into the Wilson loop superconnection \eqref{eq:cL}, and splitting the $\omega$-bilinears into $SU(3)$ irreducible representations, we find $\omega_i\bar\omega^j = [\omega_i\bar\omega^j]^T + \frac{\delta_i^j}{3}\omega_k\bar\omega^k$, where $[\omega_i\bar\omega^j]^T$ is the traceless part. The resulting expression for $\cal L$ is given by the classical $1/2$ BPS configuration $\cL_{1/2}$ \cite{Drukker:2009hy}, plus linear and quadratic terms in the fluctuations that can be organized as 
\begin{equation}
\label{eq:Lexpansion}
    \cL = \cL_{1/2} + \omega_i \mathbb{O}^i + \bar\omega^i   \bar{\mathbb{O}}_i + \omega_i\bar\omega^i \left(S_1 +\frac{1}{3}S_2 + \frac{1}{2}S_3\right) + [\omega_i\bar\omega^j]^T T_j{}^i\,.
\end{equation}
This procedure provides a direct map between fluctuations of the Wilson loop couplings and fluctuations of the string worldsheet fields in $\mathbb{CP}^3$, allowing for the identification of the corresponding symmetry channels in the defect theory. In particular, the singlet bilinear $\omega_i\bar{\omega}^i$ is naturally associated with the $SU(3)$-invariant sector generated by the operators in \eqref{eq:singlets}, while the adjoint part of the bilinear corresponds to the deformation operators in \eqref{eq:traceless}. However, we do not claim that $\omega_i\bar{\omega}^i$ is dual to the specific linear combination $S_1+\frac{1}{3}S_2+\frac{1}{2}S_3$ appearing in \eqref{eq:Lexpansion}. Beyond leading order, operator mixing is expected, and therefore $\omega_i\bar{\omega}^i$ should be identified more generally with the lightest operator of definite scaling dimension in the singlet sector.

At weak coupling, several $SU(3)$-invariant marginally irrelevant deformations exist. In the strong-coupling description these should be realized as suitable singlet composite operators built from the worldsheet fields. We propose that the bilinear $\omega_i\bar{\omega}^i$ corresponds to the lightest operator in this singlet sector. Other deformations are expected to correspond to higher-order singlet operators constructed from $\omega_i$ and possibly mixing with operators involving worldsheet fermions.
Given that there is no direct identification between the weak-coupling operators $S_i$ and individual string fluctuations at strong coupling, it is also difficult to determine how the absence of an anomalous dimension for $S_1$ at weak coupling is reflected, or eventually lifted, in the string regime.

At strong coupling the classical scaling dimension of the corresponding defect operator is $\Delta = 2$, implying that it generates an irrelevant deformation. This behavior is consistent with the weak-coupling picture, where the same deformation is marginally irrelevant.

\paragraph{The anomalous dimension of $\omega_i\bar\omega^i$.}
The anomalous dimension of this operator can be extracted using the results of \cite{Bianchi:2020hsz}. In that work, the four-point function $\left\langle \omega_i(\tau_1)\bar\omega^j(\tau_2)\omega_k(\tau_3)\bar\omega^l(\tau_4) \right\rangle$ was determined both by bootstrap methods\footnote{Analytic bootstrap methods have also been applied to other Chern-Simons-matter theories in \cite{Pozzi:2024xnu, Pozzi:2025goj}.} and by an explicit computation using AdS$_2$ Witten diagrams derived from the worldsheet action \eqref{eq:Nambu-Goto1/2}. 

Conformal invariance of the one-dimensional defect strongly constrains the form of the correlator to
\begin{equation}\label{eq:4pt}
    \left\langle \omega_i(\tau_1) \bar\omega^j(\tau_2) \omega_k(\tau_3)\bar\omega^l(\tau_4) \right\rangle = \frac{C_{\omega}(\lambda)^2}{\tau_{12}^2 \tau_{34}^2} G_{ik}^{jl}(\chi)\,,
\end{equation}
where $\chi=\dfrac{\tau_{12}\tau_{34}}{\tau_{13}\tau_{24}}$, with $\tau_{mn}=\tau_m-\tau_n$, is the conformally invariant cross-ratio and $G_{ik}^{jl}(\chi)$ admits an expansion in terms of conformal blocks, as reviewed below.\footnote{\textcolor{blue} The symbol $\chi$ is not to be confused with the couplings $\chi_i$ in \eqref{eq:beta}.} We decompose the correlator according to the $SU(3)$ irreducible representations appearing in the operator product expansion, namely the singlet ($G_S$) and the traceless ($G_T$) channels
\begin{equation}\label{eq:sta}
    G_{ik}^{jl}(\chi) = G_S \delta_i^j\delta_k^l + G_T\left( \delta_i^l\delta_k^j - \frac{1}{3}\delta_i^j\delta_k^l \right)\,.
\end{equation}
At strong coupling, both functions can be expanded in power series as 
\begin{equation}
    G_{S,T}(\chi) = G_{S,T}^{(0)}(\chi) + \frac{1}{\sqrt{\lambda}} G^{(1)}_{S,T}(\chi) + \cO\left( 
        \frac{1}{\lambda} \right)\,.
\end{equation}
These coefficients can be easily evaluated by reading the singlet and the traceless contributions in the calculation of \cite{Bianchi:2020hsz}. 
At tree level,
\begin{equation}\label{eq:disconn_res}
    G_S^{(0)} =1 + \frac{1}{3}\frac{\chi^2}{(1-\chi)^2}\,, \qquad G_T^{(0)} = \frac{\chi^2}{(1-\chi)^2} \,,
\end{equation}
while the one-loop coefficients read 
\begin{equation}
\label{eq:GSGT}
\begin{split}
    G_S^{(1)}(\chi)& = -\frac{5}{6}\chi^2 -\chi^2\log\chi +\cO(\chi^3)\,,\\ G_T^{(1)}(\chi)&= -\frac{9}{4}\chi^2-\frac32 \chi^2\log\chi\, +\cO(\chi^3)\,,
\end{split}
\end{equation}
where we have already performed a series expansion in the cross ratio, in view of the OPE limit $\chi\to 0$. The terms proportional to $\chi^2\log\chi$ in these expressions signal the presence of non-vanishing anomalous dimensions for the singlet and traceless two-particle operators $\omega_i\bar\omega^i$ and $[\omega_{i}\bar\omega^{j}]^T$.

To extract these anomalous dimensions we consider the OPE limit of the four-point function, in particular the expansion of the conformal block for the four-point function of a complex scalar field, which reads
\begin{equation}\label{eq:ope}
   G (z) = 1+\sum_{\Delta} c_{\Delta}(-z)^{\Delta} \pFq{2}{1}{\Delta,\Delta}{2\Delta+3}{z} \equiv 1+\sum c_{\Delta} G_{\Delta}(z),\qquad z=\frac{\chi}{\chi-1}\,,
\end{equation}
where $c_{\Delta}$ are the OPE coefficients and the sum runs over the scaling dimensions $\Delta>0$ of the exchanged primary operators. At leading order, the anomalous dimensions $\gamma^{(1)}_n$ of the singlet operator and its derivatives can then be extracted from the coefficient of the $\log\chi$ term in the small-$\chi$ expansion, yielding
\begin{equation}\label{eq:gs-log}
    G_S^{(1)}\big|_{\log\chi} = \sum_{n=0}^{\infty} c_n^{(0)}\gamma_n^{(1)}G_{2+n}(z)\,.
\end{equation}
We are interested in the anomalous dimension of the lightest singlet operator, corresponding to $n=0$. The associated leading-order OPE coefficient $c_0^{(0)}$ can be determined by comparing \eqref{eq:ope} with the contribution \eqref{eq:disconn_res}, yielding $c_0^{(0)}=1/3$.

At this order no operator mixing is expected for the lightest state. Therefore, exploiting the orthogonality properties of the functions $G_\Delta$ \cite{Heemskerk:2009pn}, the relation \eqref{eq:gs-log} can be inverted to obtain
\begin{equation}\label{eq:gamma1}
    \gamma_n^{(1)} = \frac{1}{c_n^{(0)}} \oint \frac{d z}{2\pi i}\frac{z}{(1-z)^3} G_{-4-n}(z) 
   \left( G_S^{(1)}(z) \big|_{\log z}\right)\,.
\end{equation}
Substituting into \eqref{eq:gamma1} the logarithmic part of $G_S^{(1)}$ as extracted from \eqref{eq:GSGT}, and using the explicit expression $G_{-4}(z)=\frac{(z-5)(z-1)^3}{5z^4}$, the integral above can be evaluated straightforwardly. This yields $\gamma^{(1)}_{\omega_i\bar\omega^i} = -3$. Combining this result with the classical scaling dimension at strong coupling, we conclude that, at first order, the singlet operator $\omega_i\bar\omega^i$ has dimension 
\begin{equation}
\label{eq:Deltasinglet}
\Delta_{\omega_i\bar\omega^i} = 2-\frac{3}{\sqrt{\lambda}} +\mathcal{O}\left(\frac{1}{\lambda}\right)\, .
\end{equation}
This result is consistent with the perturbative analysis indicating that $W_{1/2}^+$ sits at an IR-stable fixed point \cite{Castiglioni:2022yes,Castiglioni:2023uus}. Moreover, the negative sign of the leading correction suggests a smooth interpolation toward a marginally irrelevant deformation at weak coupling.

As a side remark, the same analysis can be repeated for the traceless channel using $G_T^{(1)}$ in \eqref{eq:GSGT}, allowing to extract the anomalous dimension of the operator $[\omega_i\bar\omega^j]^T$. In this case the leading-order OPE coefficient is $c_0^{(0)}=1$. Evaluating the integral in \eqref{eq:gamma1}, with $G_S^{(1)}$ replaced by $G_T^{(1)}$, yields
\begin{equation}
\label{deltaomegaomegatraceless}
\Delta_{[\omega_i\bar\omega^j]^T} = 2- \frac{3}{2\sqrt{\lambda}} +\mathcal{O}\left(\frac{1}{\lambda}\right)\,.
\end{equation}
Since the operator $T_i{}^{j}$ is already irrelevant at the classical level, this result further corroborates the stability of the IR fixed point.
\section{The bosonic \texorpdfstring{$1/6$ BPS loop $W_{1/6}$}{1/6 BPS loop W1/6}}
\label{sec:W16}
\vskip 7pt

We now turn to the $1/6$ BPS bosonic Wilson loop. As recalled in section \ref{sec:general} and illustrated in figure \ref{fig:fluxes}, at weak coupling this defect can play two different roles in the space of defect RG flows. On the one hand, it arises as the IR fixed point of flows originating from UV theories, most notably the $W^{-}$ fixed point. On the other hand, it admits relevant deformations that drive the flow toward other IR fixed points, such as $W_{1/2}^{+}$ and $W^{+}$. 

Particularly interesting is the so-called enriched flow connecting $W_{1/6}$ and $W_{1/2}^{+}$ along a trajectory that preserves a residual amount of supersymmetry, even if it does not correspond to a conformal defect at the quantum level \cite{Castiglioni:2022yes}. Another trajectory, driven by a purely bosonic deformation, connects $W^{-}$ to $W_{1/6}$ while preserving an $SU(2)\times SU(2)$ subgroup of the $SU(4)$ R-symmetry that is manifest at the $W^{-}$ point.

In this section we investigate the string dual description of the $1/6$ BPS operator and search for the corresponding deformations that induce RG flows at strong coupling. We use the $SU(2)\times SU(2)$ symmetry of the defect as a guiding principle to identify the string fluctuations corresponding to operators in the weakly coupled field theory description, and study their scaling dimensions by computing their leading corrections at strong coupling. If a picture analogous to the weak-coupling analysis persists in this regime, we expect to find both relevant and irrelevant deformations of the defect fixed point.

The $W_{1/6}$ operator is dual to an AdS$_2$ string solution smeared over a $\mathbb{CP}^1 \subset \mathbb{CP}^3$ \cite{Drukker:2008zx}. In this case, the $\mathbb{CP}^3$ sector of the string action must be treated appropriately in order to preserve the $SU(2)\times SU(2)$ internal symmetry characteristic of the $W_{1/6}$ configuration. To this end, it is convenient to parametrize $\mathbb{CP}^3$ using the Hopf coordinates in \eqref{eq:hopf}.

We begin by imposing the Dirichlet boundary condition  $\alpha=0$, in order to reduce the $\mathbb{CP}^3$ metric \eqref{eq:cp3angular} to 
\begin{equation}
    ds^2 = \frac{1}{4}\left( d\theta_1^2 + \sin^2\theta_1d \varphi_1^2 \right)\,,
\end{equation}
which describes the $\mathbb{CP}^1\simeq S^2$ subset of $\mathbb{CP}^3$ on which the string is smeared. In homogeneous coordinates, this translates into
\begin{equation}
\label{eq:bc16}
    Z_1 = \cos\frac{\theta_1}{2}e^{i\frac{\varphi_1}{2}}\,,\qquad
    Z_2 = \sin\frac{\theta_1}{2}e^{-i\frac{\varphi_1}{2}}\,,\qquad
    Z_3  = 0\,,\qquad
    Z_4  = 0\, . 
\end{equation}
We then impose Neumann boundary conditions  $\partial_\sigma\theta_1|_{\sigma = 0} = \partial_\sigma\varphi_1|_{\sigma = 0}=0$. As a result, the system exhibits $SU(2)_1 \times SU(2)_2$ symmetry. It is therefore convenient to split the $Z_I$ 4-vector as
\begin{equation}
    Z_I = (z_i, w_{\hat \imath})\,,\qquad 
    \bar Z^I= (\bar z^i, \bar w^{\hat \imath})\, ,
\end{equation}
where $i =1,2$ ($\hat\imath=3,4$) are $SU(2)_1$ ($SU(2)_2$) indices. According to \eqref{eq:bc16}, we must impose Neumann boundary conditions on $z_i$ and Dirichlet ones on $w_{\hat \imath}$.

The classical solution is given by $Z_I = (n_i, 0), \, \bar{Z}^I = (\bar{n}^i, 0)$, where $n_i$ are constant 2-vectors of $SU(2)_1$ such that $n_i \bar{n}^i = 1$, integrated over $\mathbb{CP}^1$. Transverse quantum fluctuations can be naturally introduced by the following parametrization
\begin{equation}
\label{eq:fluctuations}
\begin{split}
    %&
    Z_I = \left(\sqrt{1-\frac{\zeta^2}{2} -\frac{w^2}{2}} \,n_i + \frac{\zeta_i}{\sqrt{2}}, \frac{w_{\hat\imath}}{\sqrt{2}}\right) \,, \qquad
    \bar{Z}^I = \left(\sqrt{1-\frac{\zeta^2}{2} -\frac{w^2}{2}} \,\bar{n}^i + \frac{\bar\zeta^i}{\sqrt{2}}, \frac{\bar{w}^{\hat\imath}}{\sqrt{2}}\right) \,, 
    \end{split}
\end{equation}
with 
\begin{equation}
\label{eq:fluctconstraint}
\zeta_i \bar n^i = n_i \bar \zeta^i = 0.
\end{equation}
In the expressions above, $\zeta^2 = \zeta_i \bar \zeta^i$ and $w^2 = w_{\hat \imath} \bar{w}^{\hat \imath}$ are small functions and suitable factors have been introduced to obtain canonically normalized kinetic terms in the following expansion. 

In terms of \eqref{eq:fluctuations}, the Nambu-Goto action takes the form
\begin{equation}\label{eq:Nambu-Goto1/6}
    S = T \int d\sigma d\tau \sqrt{\det\left[ g_{\mu\nu}+ 4\left( \frac{\partial_{\mu} \zeta_i \partial_{\nu}\bar{ \zeta}^i + \partial_{\mu} w_{\hat\imath} \partial_{\nu} \bar{w}^{\hat\imath}}{1+ |\zeta|^2 + |w|^2} - \frac{(\partial_{\mu}\bar{\zeta}^i \zeta_i+\partial_{\mu}\bar{w}^{\hat\imath}w_{\hat\imath})(\bar{\zeta}^j\partial_\nu \zeta_j +\bar{w}^{\hat\jmath}\partial_\nu w_{\hat\jmath})
    }{(1+ |\zeta|^2 + |w|^2)^2} 
    \right)\right]}\, ,
\end{equation}
where $g_{\mu\nu}$ is the AdS$_2$ metric, see  \eqref{eq:AdS2solution}. Expanding the action in powers of the fluctuation fields $\zeta_i$ and $w_{\hat\imath}$, we obtain
\begin{equation}
    S=T\int d\sigma d\tau \sqrt{g}\,(1+L_B)\,, \quad L_B=L_2+L_{4\zeta}+L_{4w}+L_{2\zeta,2w}+ \cdots\,,
\end{equation}
where
\begin{align}
\begin{split}\label{eq:bosonicexp}
    L_2&= g^{\mu\nu} \partial_\mu w_{\hat\imath}\partial_\nu\bar{w}^{\hat\imath} + g^{\mu\nu}\partial_\mu \zeta_i \partial_{\nu}\bar\zeta^i, \\
    L_{4\zeta}&=  -\frac{1}{2}|\zeta|^2(g^{\mu\nu}\partial_\mu\zeta_i\partial_\nu\bar{\zeta}^i)-\frac{1}{2}(\zeta_i\bar\zeta^j)(g^{\mu\nu}\partial_\mu\bar{\zeta}^i\partial_\nu\zeta_j)+\frac{1}{2}(g^{\mu\nu}\partial_\mu\zeta_i\partial_\nu\bar{\zeta}^i)(g^{\rho\kappa}\partial_\rho\zeta_j\partial_\kappa\bar{\zeta}^j)\\
    &\quad-\frac{1}{2}(g^{\mu\nu}\partial_\mu\zeta_i\partial_\nu\bar\zeta^j)(g^{\rho\kappa}\partial_\rho\bar\zeta^i\partial_\kappa\zeta_j)-\frac{1}{2}(g^{\mu\nu}\partial_\mu\zeta_i\partial_\nu\zeta_j)(g^{\rho\kappa}\partial_\rho\bar\zeta^i\partial_\kappa\bar\zeta^j), \\
    L_{4w}&=-\frac{1}{2}|w|^2(g^{\mu\nu} \partial_\mu w_{\hat\imath}\partial_\nu\bar{w}^{\hat\imath})-\frac{1}{2}(w_{\hat\imath}\bar{w}^{\hat\jmath})(g^{\mu\nu}\partial_\mu\bar{w}^{\hat\imath} \partial_\nu w_{\hat\jmath})+\frac{1}{2}(g^{\mu\nu}\partial_\mu w_{\hat\imath}\partial_\nu\bar{w}^{\hat\imath})(g^{\rho\kappa}\partial_\rho w_{\hat\imath}\partial_\kappa\bar{w}^{\hat\imath})\\
    &\quad -\frac{1}{2}(g^{\mu\nu}\partial_\mu w_{\hat\imath}\partial_\nu\bar{w}^{\hat\jmath})(g^{\rho\kappa}\partial_\rho\bar{w}^{\hat\imath}\partial_\kappa w_{\hat\jmath})-\frac{1}{2}(g^{\mu\nu}\partial_\mu{w}_{\hat\imath}\partial_\nu{w}_{\hat\jmath})(g^{\rho\kappa}\partial_\rho\bar{w}^{\hat\imath}\partial_\kappa\bar{w}^{\hat\jmath}), \\
    L_{2\zeta,2w}&= -\frac{1}{2}|\zeta|^2(g^{\mu\nu}\partial_\mu w_{\hat\imath}\partial_\nu\bar{w}^{\hat\imath}) - \frac{1}{2}|w|^2 (g^{\mu\nu}\partial_\mu \zeta_{i}\partial_\nu\bar{\zeta}^i)-\frac{1}{2}(\zeta_i\bar{w}^{\hat\jmath})(g^{\mu\nu}\partial_\mu\bar\zeta^i \partial_\nu w_{\hat\jmath}) \\
    &\quad-\frac{1}{2} (w_{\hat\imath}\bar\zeta^i)(g^{\mu\nu}\partial_\mu\bar{w}^{\hat\imath} \partial_\nu \zeta_i)+ (g^{\mu\nu}\partial_\mu \zeta_i\partial_\nu\bar{\zeta}^i)(g^{\rho\kappa}\partial_\rho w_{\hat\imath}\partial_\kappa\bar{w}^{\hat\imath})-(g^{\mu\nu}\partial_\mu \zeta_i\partial_\nu\bar{w}^{\hat\imath})(g^{\rho\kappa}\partial_\rho\bar\zeta^i\partial_\kappa w_{\hat\imath})\\
    &\quad -(g^{\mu\nu}\partial_\mu \zeta_i\partial_\nu w_{\hat\imath}) (g^{\rho\kappa}\partial_\rho\bar\zeta^i\partial_\kappa\bar{w}^{\hat\imath})\,.
\end{split}
\end{align}

This action describes an interacting theory of two massless fields, $\zeta_i$ transforming in the $(\mathbf{2},\mathbf{1})$ and $w_{\hat\imath}$ in the $(\mathbf{1},\mathbf{2})$ representation of $SU(2)_1\times SU(2)_2$, subject to the constraint \eqref{eq:fluctconstraint}. Composite fields can be constructed by considering products of the $z_i$ and $w_{\hat\imath}$ fields. In table \ref{tab:AdSoperators} we list the lowest-dimensional operators obtained in this way, together with their scaling dimensions and their $SU(2)_1\times SU(2)_2$ representations. Since $z_i$ and $w_{\hat\imath}$ obey different boundary conditions, they have different scaling dimensions $\Delta$. Consequently, in the strong coupling regime they are dual to defect operators with different classical dimensions.

\begin{table}[ht!]
\begin{center}
\begin{tabular}{ |c|c|c|c| } 
 \hline
 Field & $\Delta$ & $SU(2)_1$ & $SU(2)_2$ \\ 
 \hline\hline
 $z_i$ & 0& $\mathbf{2}$ & $\mathbf{1}$ \\ 
 $\bar{z}^i$ & 0& $\mathbf{\bar 2}$ & $\mathbf{1}$ \\ 
 $w_{\hat\imath}$ &1  &$\mathbf{1}$ & $\mathbf{2}$ \\ 
 $\bar w^{\hat\imath}$ &1  & $\mathbf{1}$ & $\mathbf{\bar 2}$ \\ 
 $z_i\bar{z}^i$ &0  & $\mathbf{1}$ & $\mathbf{1}$ \\ 
 $ w_{\hat\imath}\bar w^{\hat\imath} $ & 2 & $\mathbf{1}$ & $\mathbf{1}$ \\ 
 $z_i\bar w^{\hat\imath}$ & 1 & $\mathbf{2}$ & $\mathbf{\bar 2}$ \\ 
 $ w_{\hat\imath}\bar{z}^{i} $ & 1  & $\mathbf{\bar 2}$ & $\mathbf{2}$ \\ 
 \hline
\end{tabular}
\caption{\label{tab:AdSoperators} Data of lowest dimensional fields on the string theory side.}
\end{center}
\end{table}

%%%%%%%%%%%%%%%%

\subsection{Defect operators and their dual two-particle states}
\label{sec:W1/6operators}
\vskip 7pt

On the field theory side, the $1/6$ BPS Wilson loop supported along the circle is defined as
\begin{equation}
    W_{1/6} = \Tr\cP \exp\oint_\mathcal{C} d\tau\, \left( A_\mu \dot{x}^\mu - \frac{2\pi i}{k}\vert \dot x\vert M_J{}^I C_I \bar{C}^J \right)\,, 
    \qquad 
    M_J{}^I=\text{diag}(-1,-1,1,1)\,.
\end{equation}
For this defect one can identify bosonic unprotected operators, $C_i \bar{C}^i$ and $C_{\hat\imath} \bar{C}^{\hat\imath}$, as well as protected operators \cite{Bianchi:2018scb}
\begin{equation}
    \cO^{i\hat\imath}= -\frac{4\pi i}{k} \epsilon^{ij} C_j \bar{C}^{\hat\imath}\,, 
    \qquad 
    \bar\cO^{\hat{\imath} i}=-\frac{4\pi i}{k} \epsilon^{\hat\imath\hat\jmath} C_{\hat\jmath} \bar{C}^i\,.
\end{equation}
These operators are summarized in table~\ref{tab:CFToperators}, together with their quantum numbers.

\begin{table}[h!]
\begin{center}
\begin{tabular}{ |c|c|c|c| } 
 \hline
 Operator & $\Delta$ &  $SU(2)_1$ & $SU(2)_2$ \\ 
 \hline\hline
 $\cO^{i \hat\imath}$ & 1  & $\mathbf{2}$ & $\mathbf{\bar 2}$ \\ 
 $\bar\cO^{\hat\imath i}$ & 1  & $\mathbf{\bar 2}$ & $\mathbf{2}$ \\
  \hline
 $ C_i \bar C^i $ & 1 & $\mathbf{1}$ & $\mathbf{1}$ \\ 
 $ C_{\hat\imath}\bar C^{\hat\imath}$ & 1  & $\mathbf{1}$ & $\mathbf{1}$ \\
 \hline
\end{tabular}
\caption{\label{tab:CFToperators} Data of protected and unprotected operators on the CFT side.}
\end{center}
\end{table}

Naturally, a comparison between tables \ref{tab:AdSoperators} and \ref{tab:CFToperators} suggests the following identifications
\begin{equation}
 \cO^{i \hat\imath} \leftrightarrow \epsilon^{ij} z_j \bar w^{\hat\imath} \, , \qquad 
 \bar{\cO}^{\hat\imath i} \leftrightarrow \epsilon^{\hat\imath \hat\jmath} w_{\hat\jmath} \bar{z}^i \, , \qquad 
 C_i \bar{C}^i \leftrightarrow z_i \bar{z}^i \, , \qquad 
 C_{\hat\imath} \bar{C}^{\hat\imath} \leftrightarrow w_{\hat\imath} \bar{w}^{\hat\imath} \, .
\end{equation}
To support this correspondence, we compute below the anomalous dimensions of the composite fields $z_i \bar{z}^i$ and $w_{\hat\imath} \bar{w}^{\hat\imath}$ in the string dual description, verifying that these operators are indeed not protected. A direct verification that $z_i \bar{w}^{\hat\imath}$ and $w_{\hat\imath} \bar{z}^i$ are protected against acquiring an anomalous dimension would instead require a full superstring computation, which we do not attempt here. Instead, we assume this property to hold and use it to streamline the determination of the anomalous dimensions of the elementary fields $z_i$ and $w_{\hat\imath}$, which will play a central role in the subsequent analysis.

At leading order, the anomalous dimensions of $z_i \bar{w}^{\hat\imath}$ and $ w_{\hat\imath} \bar{z}^i$ are simply given by the sum of the anomalous dimensions of the elementary fields. This occurs because the four-point function $\left\langle z_i \bar w^{\hat\imath} w_{\hat \jmath} \bar z^j \right\rangle$ does not receive one-loop corrections from connected diagrams. Indeed, using the expansion $z_i \simeq n_i + \frac{\zeta_i}{\sqrt{2}}$, it is easy to see that the correlator $\left\langle z_i \bar w^{\hat\imath} w_{\hat \jmath} \bar z^j \right\rangle$ receives its leading contribution from the disconnected contraction $\left\langle n_i \bar n^j\right\rangle \left\langle w^{\hat\imath} w_{\hat \jmath}\right\rangle$, which is of order $\frac{1}{\sqrt{\lambda}}$. The first connected contribution involves instead a quartic interaction vertex together with four propagators and therefore it appears only at order $\frac{1}{\lambda^{3/2}}$, corresponding to a two-loop correction.

Therefore, to first order the anomalous dimension of $z_i \bar w^{\hat{\imath}}$ is simply the sum of the anomalous dimensions of $z_i$ and $w_{\hat\imath}$. Since this is identified with the dual of the protected defect operator ${\cal O}^{i\hat\imath}$, it follows that the anomalous dimensions of $z_i$ and $w_{\hat\imath}$ must cancel at leading order. We now compute these anomalous dimensions, before turning to the unprotected operators discussed above.

\paragraph{Anomalous dimension of $z_i$.} We extract this anomalous dimension directly from  the two-point function $\left\langle z_i(\tau_1) \bar z^j(\tau_2) \right\rangle$, with $z_i, \bar{z}^j$ expressed in terms of the quantum fluctuations $\zeta_i, \bar\zeta^{j}$ in \eqref{eq:fluctuations}. 

The propagator satisfying Neumann boundary conditions in AdS$_2$ is given by \cite{Beccaria:2019dws}
\begin{equation}\label{eq:2ptzeta}
    \left\langle \zeta_i (\tau_1) \bar \zeta^j (\tau_2) \right\rangle = -\frac{1}{2\pi}P_i^{\ j}\, \log\left(\tau_{12}^2\right) \,, \qquad \qquad P_i^{\ j} = \delta_i^j - n_i \bar n^j \,.
\end{equation}
Thus, at leading order in $\zeta_i$, the correlation function reads
\begin{equation}
\begin{split}
    \left\langle z_i(\tau_1) \bar z^j(\tau_2) \right\rangle =& \left\langle n_i \bar n^j \right\rangle + \frac{1}{2T}\left\langle \zeta_i(\tau_1) \bar \zeta^j (\tau_2) \right\rangle = \frac{\delta_i^j}{2}\left( 1 - \frac{1}{2\sqrt{\lambda}}\log\tau_{12}^2\right)\,,
\end{split}
\end{equation}
where we have used the normalization $\left\langle n_i \bar n^j \right\rangle = \frac{\delta_i^j}{2}$ and $\left\langle P_i^{\ j} \right\rangle= \frac{\delta_i^j}{2}$.

The coefficient of the logarithmic term corresponds to minus the anomalous dimension $\gamma_z$. Therefore, the operator is relevant and, at first order, we find
\begin{equation}\label{eq:Deltaz}
    \Delta_z = \frac{1}{2\sqrt{\lambda}} + {\cal O}\left( \frac{1}{\lambda} \right)\,.
\end{equation}

%%%%%%%%%%%%%%%%%%%%%%%%%%

\paragraph{Anomalous dimension of $w_{\hat \imath}$.} 
A direct determination of the anomalous dimension of $w_{\hat \imath}$ involves the calculation of loop corrections to the two-point function $\left\langle w_{\hat\imath} \bar{w}^{\hat\jmath} \right\rangle$, which possesses the tree level expression
\begin{equation}
    \left\langle w_{\hat\imath}(\tau_1) \bar w^{\hat \jmath}(\tau_2) \right\rangle = %\frac{2}{\sqrt{\lambda}} 
    \mathcal{C}_1 \frac{\delta_{\hat\imath}^{\hat \jmath}}{\tau_{12}^2}\,,
\end{equation}
with $\mathcal{C}_1=\frac{1}{\pi}$.
At one loop this receives tadpole corrections from loops of both $w_{\hat\imath}$ and $\zeta_i$ scalars from the quartic vertices in \eqref{eq:bosonicexp}. In addition, one-loop diagrams with worldsheet fermion loops would also contribute. However, their evaluation would require considering the full Green-Schwarz superstring action, which is beyond the scope of this work. Here we limit our analysis to bosonic fluctuations corrections, obtained from the quartic vertices $L_{4w}$ and $L_{2\zeta,2w}$ given in \eqref{eq:bosonicexp}.

We start with the $L_{2\zeta,2w}$ vertex, needed to compute 
\begin{equation}\label{eq:wwS}
    \left\langle w_{\hat\imath}(\tau_1) \bar w^{\hat \jmath}(\tau_2) \right\rangle^{(1)} \bigg|_\zeta = -\left\langle w_{\hat\imath}(\tau_1) \bar w^{\hat \jmath}(\tau_2)\int L_{2\zeta,2w} \right\rangle\,.
\end{equation}
This diagram requires the Dirichlet bulk-to-boundary propagator for $w_{\hat\imath}$ 
\begin{equation}\label{eq:bulktoboundaryw}
    K(z,\tau;\tau_n)\equiv K_n(z,\tau) = \mathcal{C}_1 \frac{z}{(\tau_n-\tau)^2+z^2} \,,
\end{equation}
and the bulk-to-bulk AdS$_2$ propagator for a field with Neumann boundary conditions like $\zeta_i$, taken in the limit of coincident points because of the tadpole diagram
\begin{equation}
     N(z,\tau) =  - \frac{1}{4\pi}\, \lim_{(z',\tau')\to (z,\tau)} \Big( \log\left[(\tau-\tau') + (z-z')^2 +\epsilon^2 z z'  \right] + \log\left[ (\tau-\tau')^2 +(z+z')^2 \right] \Big) \,.
\end{equation}
We obtain
\begin{equation}\label{eq:ww_1loop}
\begin{split}
     \left\langle w_{\hat\imath}(\tau_1) \bar w^{\hat \jmath}(\tau_2) \right\rangle^{(1)} \,\bigg|_\zeta &
     = \frac{\delta_{\hat \imath}^{\hat\jmath}}{T}\int \frac{dzd\tau}{z^2} \,  g^{\mu\nu} \bigg(
     \partial_\mu K_1 \partial_\nu K_2 N + 
     \partial_\mu (K_1 K_2) \partial_\nu N +
     K_1 K_2 \partial_\mu\partial_\nu N\\&
     \hskip 3cm +4 \partial_\mu K_1 \partial^\rho K_2 \partial_\nu \partial_\rho N
     -2 \partial_\mu K_1 \partial_\nu K_2 \partial^2 N
 \bigg)\,.
\end{split}
\end{equation}

Focusing only on the terms proportional to $\log\tau_{12}^2$ arising from the integrals in \eqref{eq:ww_1loop}, and discarding linear divergences, we find the following non-vanishing contributions
\begin{equation}
\begin{split}
    \int \frac{dzd\tau}{z^2} g^{\mu\nu} K_1K_2 \partial_{\mu}\partial_{\nu}N \,\bigg|_\text{log} &= 
    -\frac{\mathbf{k}}{4\pi^2} \frac{\log \tau_{12}^2}{\tau_{12}^2}\,, \\
    \int \frac{dzd\tau}{z^2} g^{\mu\nu} \partial_{\mu} (K_1 K_2) \partial_{\nu} N\,\bigg|_\text{log} &= \frac{1}{2\pi^2} \frac{\log \tau_{12}^2}{\tau_{12}^2}\,,\\
    \int \frac{dzd\tau}{z^2} g^{\mu\nu} \partial_{\mu} K_1 \partial_{\nu}K_2 N\,\bigg|_\text{log} &= -\frac{1}{2\pi^2} \frac{\log \tau_{12}^2}{\tau_{12}^2}\,.
\end{split}
\end{equation}
The constant $\mathbf{k}$ encodes the regularization scheme dependence of the propagator. For consistency with the string theory symmetries, in AdS$_2$ it is typically chosen to be equal to 2 \cite{Tseytlin:1988ne}, but for now we leave it arbitrary. Summing up everything, we find that the $\log\tau_{12}^2$ term appearing in the one-loop correction of the two-point function of $w_{\hat\imath}$ is given by
\begin{equation}
\begin{split}
    \left\langle w_{\hat\imath}(\tau_1) \bar w^{\hat \jmath}(\tau_2) \right\rangle^{(1)} \,\bigg|_\zeta =
    -\mathcal{C}_1 \frac{\delta_{\hat\imath}^{\hat \jmath}}{\tau_{12}^2}
    %\left( 1+ 
    \frac{1}{\sqrt{\lambda}} \frac{\mathbf{k}}{2} \log\tau_{12}^2 %%\right)
    \,.
\end{split}
\end{equation}

Now we consider the contribution from the $L_{4w}$ vertex
\begin{equation}
\left\langle w_{\hat\imath}(\tau_1) \bar w^{\hat \jmath}(\tau_2) \right\rangle^{(1)} \bigg|_w = -\left\langle w_{\hat\imath}(\tau_1) \bar w^{\hat \jmath}(\tau_2)\int \! L_{4w} \right\rangle,
\end{equation}
which requires the Dirichlet bulk-to-boundary propagator for $w_{\hat\imath}$ \eqref{eq:bulktoboundaryw}, as well as its bulk-to-bulk propagator
\begin{equation}\label{eq:bulktobulkw}
    \textrm{D}(\tau,z;\tau',z')=-\frac{1}{4\pi}\bigg(\log\left[(\tau-\tau')^2+(z-z')^2+\epsilon^2 zz'\right]-\log\left[(\tau-\tau')^2+(z+z')^2\right]\bigg)\,.
\end{equation}
We obtain
\begin{align}
   \label{eq:ww_1loop_4w}
     \left\langle w_{\hat\imath}(\tau_1) \bar w^{\hat \jmath}(\tau_2) \right\rangle^{(1)} \bigg|_w &=
\frac{\delta_{\hat\imath}^{\hat\jmath}}{T}\int \frac{dzd\tau}{z^2} \,g^{\mu\nu} \bigg( \frac{3}{2}
     \partial_\mu K_1 \partial_\nu K_2 \textrm{D} + \frac{3}{2}
     \partial_\mu (K_1 K_2) \partial_\nu \textrm{D} + \frac{3}{2}
     K_1 K_2 \partial_\mu\partial_\nu \textrm{D}
\cr &\hskip 3.5cm     - \partial_\mu K_1 \partial_\nu K_2 \partial^2 \textrm{D}
     +4 \partial^\rho K_1 \partial_\mu K_2 \partial_\nu\partial_\rho \textrm{D}
 \bigg).
\end{align}
Only the following integral contributes to logarithmic terms
\begin{equation}
    \int \frac{dzd\tau}{z^2} g^{\mu\nu} K_1K_2 \partial_{\mu}\partial_{\nu}\textrm{D} \,\bigg|_\text{log}=
    -\frac{\mathbf{k}}{4\pi^2} \frac{\log \tau_{12}^2}{\tau_{12}^2}\,,
\end{equation}
so that
\begin{equation}
\begin{split}
    \left\langle w_{\hat\imath}(\tau_1) \bar w^{\hat \jmath}(\tau_2) \right\rangle^{(1)} \,\bigg|_w =
    -\mathcal{C}_1 \frac{\delta_{\hat\imath}^{\hat \jmath}}{\tau_{12}^2}
    \frac{1}{\sqrt{\lambda}} \frac{3\mathbf{k}}{4} \log\tau_{12}^2 
    \,.
\end{split}
\end{equation}
Altogether, the bosonic contributions give
\begin{equation}
\begin{split}
    \left\langle w_{\hat\imath}(\tau_1) \bar w^{\hat \jmath}(\tau_2) \right\rangle \,\bigg|_\text{bosonic} =
    \mathcal{C}_1 \frac{\delta_{\hat\imath}^{\hat \jmath}}{\tau_{12}^2}
    \left( 1-
    \frac{5\mathbf{k}}{4\sqrt{\lambda}} \log\tau_{12}^2 \right)
    \,.
\end{split}
\end{equation}
For $\mathbf{k}=2$, the bosonic contributions to the anomalous dimension are
\begin{equation}
    \gamma^{(1)}_{w}\,\Big|_\text{bosonic} = \frac{5}{2\sqrt{\lambda}}+\mathcal{O}\left(\frac{1}{\lambda}\right)\,.
\end{equation}
As discussed above, at one loop the anomalous dimension of the composite operator $z_i \bar w^{\hat \imath}$ is simply given by the sum of the anomalous dimensions of its constituent fields. According to the protected nature of the corresponding multiplet, the total anomalous dimension should vanish
\begin{equation}
    \gamma_{w} + \gamma_{z}
    =
    \mathcal{O}\!\left(\frac{1}{\lambda}\right)\,.
\end{equation}
This cancellation involves the complete anomalous dimension $\gamma_w$, including contributions from both bosonic and fermionic worldsheet fluctuations. While our analysis only determines the bosonic contribution explicitly, supersymmetry strongly constrains the structure of the full result. Assuming the expected cancellation mechanism implied by multiplet protection, we infer that the fermionic fluctuations must contribute
\begin{equation}
\label{eq:fermioniccontribution}
    \gamma^{(1)}_{w}\,\Big|_\text{fermionic} = -\frac{3}{\sqrt{\lambda}}+ \mathcal{O}\left(\frac{1}{\lambda}\right)\,.
\end{equation}
This result should therefore be regarded as an inference based on supersymmetry and multiplet protection rather than as the outcome of a direct worldsheet computation.
This partial cancelation between the bosonic and fermionic loops, which is required for this mechanism to operate, is typical of supersymmetry. A complete cancelation occurs for the fluctuation fields of the $1/2$ BPS Wilson loop, which are likewise protected by supersymmetry.

%%%%%%%%%%%%%%%%%%%%%%%%%%

\paragraph{Anomalous dimension of $z_{i}\bar z^j$.} 

We now turn to the evaluation of the anomalous dimensions of the composites $z_{i}\bar z^j$, dual to the corresponding operators $C_i\bar C^j$ in table \ref{tab:CFToperators} with classical dimension 1. Their one-loop anomalous dimension can be extracted from the four-point function $\left\langle z \bar z z \bar z \right\rangle$.

At first order, expanding $z_i\simeq  n_i+\frac{\zeta_i}{\sqrt{2}}$, only simple exchange diagrams contribute, for which we use the two-point functions \eqref{eq:2ptzeta}.
The four-point function reads
\begin{align}\label{eq:zzzz}
    \left\langle z_i (\tau_1) \bar z^j (\tau_2) z_k (\tau_3) \bar z^l (\tau_4) \right\rangle &=
     \frac{1}{6}\bigg[\delta_i^j \delta_k^l + \delta_i^l \delta_k^j
    +\frac{1}{2\sqrt{\lambda}} \Big(\delta_i^l \delta_k^j\left(\log\tau_{12}^2-2 \log\tau_{14}^2-2
   \log\tau_{23}^2+\log \tau_{34}^2\right)  \nonumber\\& \hskip 1cm
-\delta_i^j \delta_k^l \left(2
   \log \tau_{12}^2-\log \tau_{14}^2-\log \tau_{23}^2 +2 \log
   \tau_{34}^2\right)  \Big)
    \bigg]+ \mathcal{O}\left(\frac{1}{\lambda}\right)\,.
\end{align}
We further decompose the four-point function into $SU(2)_1$ singlet and adjoint components 
\begin{equation}\label{eq:decompositionzzzz}
    \left\langle z_i (\tau_1) \bar z^j (\tau_2) z_k (\tau_3) \bar z^l (\tau_4) \right\rangle = \frac{C_z^2}{\tau_{12}^{2\Delta}\tau_{34}^{2\Delta}}\, \left(G_S \delta_i^j\delta_k^l + G_T\left( \delta_i^l\delta_k^j - \frac{1}{2}\delta_i^j\delta_k^l \right)\right)\,,
\end{equation}
and expand in $\frac{1}{\sqrt{\lambda}}$, finding
\begin{equation}
    G_S = 1 + \mathcal{O}\left(\frac{1}{\lambda}\right)\, , \qquad G_T = \frac23 +\frac83 \frac{1}{\sqrt{2\lambda}} \log\frac{\chi}{\chi-1}+  \mathcal{O}\left(\frac{1}{\lambda}\right)\,.
\end{equation}
Extracting the anomalous dimensions, we obtain
\begin{equation}
\Delta_{z_i\bar z^i} = \mathcal{O}\left(\frac{1}{\lambda}\right)\, , \qquad \Delta_{[z_i\bar z^j]^T} = \frac{4}{2\sqrt{\lambda}} +  \mathcal{O}\left(\frac{1}{\lambda}\right)    \,.
\end{equation}
Both correspond to relevant defect operators, driving RG flows from $W_{1/6}$ in the UV.

%%%%%%%%%%%%%%%%%%%%%%%%%%

\paragraph{Anomalous dimension of $w_{\hat \imath}\bar w^{\hat \jmath}$.} 

According to our identifications, we determine the first order anomalous dimension of the operators $C_{\hat\imath}\bar C^{\hat\jmath}$ from the OPE analysis of the four-point function $\left\langle w \bar w w \bar w \right\rangle$. The calculation parallels that of the 1/2 BPS defect in \cite{Bianchi:2020hsz}. Compared to that case, the external fields have different indices and possess an anomalous dimension $\Delta = 1 + \frac{1}{\sqrt{\lambda}} +  \mathcal{O}\left(\frac{1}{\lambda}\right)$. Nonetheless, the non-trivial part, that is the calculation of the contact diagram, is identical, because the $L_{4w}$ vertex in \eqref{eq:bosonicexp} possesses the same structure. 

Performing the decomposition in the $SU(2)_2$ singlet and adjoint channels, similarly to what done in  \eqref{eq:decompositionzzzz}, we find
\begin{equation}
    G_S^{(0)} = 1+\frac{\chi^2}{2}\, ,\qquad G_T^{(0)} = \chi^2\,,
\end{equation}
and
\begin{align}
    & G_S^{(1)} = \frac{-9 \chi ^3+22 \chi ^2+\left(4 (\chi -1)^3-\chi +5\right) \chi ^2 \log (\chi )-21 \chi +8}{4(\chi -1)^3}+\frac{(\chi -8) \log (1-\chi )}{\chi}\,,\\ 
    & G_T^{(1)} = \frac{\chi  \left(-3 \chi ^2+2 \chi +\left(4 (\chi -1)^3+\chi +3\right) \chi  \log (\chi )+1\right)}{2(\chi -1)^3}-\frac{\log (1-\chi )}{2}\,.
\end{align}
After taking the OPE limit $\chi\to 0$, and comparing coefficients of the relevant powers of $\chi$, we find the following OPE data for the lowest dimension exchanged operators in the singlet and adjoint channels
\begin{align}
    c_S^{(0)} = \frac12, \qquad \gamma_S^{(1)} = -\frac12\frac{1}{\sqrt{\lambda}},\qquad c_T^{(0)} = 1,\qquad \gamma_T^{(1)} = \frac12\frac{1}{\sqrt{\lambda}}\,.
\end{align}
We conclude that the operators possess dimensions
\begin{equation}
\Delta_{w_{\hat{\imath}}\bar w^{\hat{\imath}}} = 2-\frac12\frac{1}{\sqrt{\lambda}}+\mathcal{O}\left(\frac{1}{\lambda}\right), \qquad \Delta_{[w_{\hat{\imath}}\bar w^{\hat{\jmath}}]^T} = 2 +\frac12\frac{1}{\sqrt{\lambda}} +  \mathcal{O}\left(\frac{1}{\lambda}\right) \,,
\end{equation}
corresponding to irrelevant deformations of the 1/6 BPS defect fixed point.

Together with the previous results, this confirms the weak-coupling picture in which the $1/6$ BPS configuration behaves as an unstable fixed point of the RG flow. In particular, it is tempting to identify the $SU(2)$ singlet deformation associated with the $w$ fields as the endpoint of the $SU(2)\times SU(2)$ symmetric flow connecting $W^{-}$ to $W_{1/6}$. It is natural that this deformation involves the $w_{\hat \imath}$ coordinates, whose boundary conditions are expected to change from Neumann in the UV to Dirichlet in the IR through the insertion of an appropriate boundary operator.
\section{The ordinary Wilson loop \texorpdfstring{$W^-$}{W-}}
\label{sec:Wminus}
\vskip 7pt

The $W^{-}$ Wilson loop was introduced in \cite{Castiglioni:2023uus} as a bosonic $SU(4)$ invariant operator that does not preserve any supersymmetry. It has been interpreted as the ABJM analogue of the ordinary Wilson loop $\Tr \cP e^{-i \oint A}$ of pure Chern-Simons theory. As in  $\mathcal{N}=4$ super Yang-Mills, it plays the role of a UV fixed point for RG flows between Wilson loops, but, in contrast to that case, it requires an additional coupling to matter fields for consistency at the quantum level.

We consider classically marginal scalar deformations of the theory, by operator insertions of the form $C_I \bar{C}^J$ with $I,J=1,2,3,4$. These are classified into the singlet $C_I \bar C^I$ and the adjoint $\left[C_I \bar C^J\right]^T$ representations of the $SU(4)$ R-symmetry group preserved by the defect. At weak coupling, all of these operators acquire anomalous dimensions and behave as marginally relevant deformations \cite{Castiglioni:2023uus}. Denoting the deformation parameter by $\upsilon$, the associated RG flows are governed by the same one-loop $\beta$-function, $\beta_{\upsilon} = \frac{N}{k}(\upsilon-1)\upsilon$, so that they share the same anomalous dimension, $\gamma^{(1)} = \left.\frac{\partial \beta_{\upsilon}}{\partial \upsilon}\right|_{\upsilon=0} = -\frac{N}{k}$. Consequently, all the operators introduced above have non-protected dimensions which read at one loop
\begin{equation}
    \Delta = 1 - \frac{N}{k} + {\cal O} \left(\frac{N^2}{k^2} \right)\,.
\end{equation}
They are all marginally relevant and drive the system away from the $W^-$ UV fixed point. 

On the string theory side, a direct counterpart of the non-supersymmetric $SU(4)$ invariant Wilson loop can be obtained by generalizing the smearing procedure employed for the string solution dual to $W_{1/6}$ in section~\ref{sec:W16}. More precisely, we impose Neumann boundary conditions in $\mathbb{CP}^3$ and smear the fundamental string solution over all $\mathbb{CP}^3$ directions. This smearing restores $SU(4)$ invariance while breaking all supersymmetries preserved by the bulk theory, furnishing an appropriate dual description of $W^-$.  

We start from the classical string solution $Z_I=n_I$, $\bar{Z}^I=\bar{n}^I$, where $n_I, \bar{n}^I$ satisfy $n_I \bar{n}^I = 1$, and consider small transverse fluctuations by setting\footnote{Here and in section \ref{sec:Wplus}, we use similar notations for the fluctuation fields as in section \ref{sec:W16}. Since the fields arise in different backgrounds and are never discussed simultaneously, no confusion should arise.} 
\begin{equation}
\begin{split}
    Z_I = \sqrt{1-\frac{\zeta^2}{2}} \,n_I + \frac{\zeta_I}{\sqrt{2}} \,, \qquad \bar Z^I = \sqrt{1-\frac{\zeta^2}{2}} \,\bar n^I + \frac{\bar\zeta^I}{\sqrt{2}} \,,
    \end{split}
\end{equation}
with
\begin{equation}
n_I \bar \zeta^I = \zeta_I \bar{n}^I = 0.
\end{equation}
Expanding in powers of the fluctuations and integrating $n_I$ over $\mathbb{CP}^3$, the string partition function reads
\begin{equation}\label{eq:partition}
{\cal Z} = \int_{\mathbb{CP}^3} dn \ d \bar n \ \delta(n_I \bar n^I -1) \int \cD\zeta \cD \bar\zeta \ \delta(n_I \bar \zeta^I) \delta(\zeta_I \bar n^I) \exp\left( -T \int d^2\sigma \sqrt{g}(L_2(\zeta) + L_4(\zeta)+\ldots)\right) \,,
\end{equation}
where
\begin{equation}
\begin{split}
    L_2(\zeta) &= \partial_\mu \zeta_I \partial^\mu \bar \zeta^I \,, \\
    L_4(\zeta) &=   -\frac{1}{2}|\zeta|^2(g^{\mu\nu}\partial_\mu\zeta_I\partial_\nu\bar{\zeta}^I)-\frac{1}{2}(\zeta_I\bar\zeta^J)(g^{\mu\nu}\partial_\mu\bar{\zeta}^I\partial_\nu\zeta_J)+\frac{1}{2}(g^{\mu\nu}\partial_\mu\zeta_I\partial_\nu\bar{\zeta}^I)(g^{\rho\kappa}\partial_\rho\zeta_J\partial_\kappa\bar{\zeta}^J)\\
    &\quad-\frac{1}{2}(g^{\mu\nu}\partial_\mu\zeta_I\partial_\nu\bar\zeta^J)(g^{\rho\kappa}\partial_\rho\bar\zeta^I\partial_\kappa\zeta_J)-\frac{1}{2}(g^{\mu\nu}\partial_\mu\zeta_I\partial_\nu\zeta_J)(g^{\rho\kappa}\partial_\rho\bar\zeta^I\partial_\kappa\bar\zeta^J)\,.
\end{split}
\end{equation}

We now consider composite bilinear fields that can correspond to the CFT operators mentioned above, using R-symmetry as a guiding criterion. This suggests identifying the bilinear operators $C_I \bar C^J$ with the field composites $Z_I \bar Z^J$. Since we imposed Neumann boundary conditions on the $\mathbb{CP}^3$ directions, the classical dimension of $Z_I$ is zero. These composites are then consistently dual to boundary relevant operators, as expected from the weak-coupling anlysis. 

In the rest of the section, we provide more details on the string dual of the defect theory, by calculating the anomalous dimensions of these fields. 

We start by considering the two-point function $\langle Z_I \bar Z^J\rangle$, to determine the first-order anomalous dimension of $Z$ directly from the quantum corrections to the propagator, as done for the $z_i$ field in section \ref{sec:W1/6operators}.  The AdS$_2$ propagator for the fluctuations with Neumann boundary condition reads \cite{Beccaria:2019dws} 
\begin{equation}
\label{eq:zeta_prop_neumann}
     \left\langle \zeta_I(\tau_1) \bar \zeta^J(\tau_2) \right\rangle = -\frac{P_I^J}{2\pi}\log \tau_{12}^2  \,, \qquad\qquad P_I^J = \delta_I^J - n_I \bar n^J\,.
\end{equation}
To perform the $\mathbb{CP}^3$ integration over the zero-modes, we choose the natural normalization $\left\langle n_I \bar n^J\right\rangle = \frac{1}{4}\delta_I^J$, whence $\left\langle P_I^J \right\rangle = \frac{3}{4}\delta_I^J$. At leading order in the fluctuations we then have
\begin{equation}
    \left\langle Z_I(\tau_1) \bar Z^J (\tau_2) \right\rangle = \left\langle  n_I \bar n^J\right\rangle + \frac{1}{2T} \left\langle \zeta_I (\tau_1) \bar \zeta^J (\tau_2) \right\rangle + \mathcal{O}\left(\frac{1}{T^2}\right) = \frac{\delta_I^J}{4}\left( 1 - \frac{3}{2\sqrt{\lambda}}\log\tau_{12}^2 \right)+  \mathcal{O}\left(\frac{1}{\lambda}\right)\,.
\end{equation}

We now consider the four-point function $\left\langle Z_I \bar Z^J Z_K \bar Z^L \right\rangle$, from which we can extract the anomalous dimensions of the bilinear composite fields $Z_I\bar Z^J$. The calculation parallels that of the four-point function \eqref{eq:zzzz}.

At first order, we expand $Z_I\simeq  n_I+\frac{1}{\sqrt{2}}\zeta_I$ and retain simple exchange diagrams, evaluated in terms of the two-point functions \eqref{eq:zeta_prop_neumann}. After decomposing the four-point function into $SU(4)$ singlet and adjoint exchanges, we find a vanishing anomalous dimension for the singlet, consistent with the exchanged operator being the identity, and 
\begin{equation}\label{eq:ZZ}
    \Delta_{[Z_I \bar Z^J]^T} = \frac{4}{\sqrt{\lambda}}+  \mathcal{O}\left(\frac{1}{\lambda}\right)\,.
\end{equation}
All $Z_I \bar Z^J$ composite fields are dual to relevant defect operator insertions, which trigger RG flows away from the UV fixed point $W^-$. 

\vskip 10pt

We can alternatively compute these anomalous dimensions by studying the spectrum of the Laplace-Beltrami operator on $ \mathbb{CP}^3$. At one loop in the sigma model, scalar operators are renormalized by an anomalous dimension operator proportional to the target-space Laplacian, $\gamma \propto \nabla^2$, so expanding them in target-space harmonics yields anomalous dimensions proportional to their Laplacian eigenvalues \cite{Tseytlin:1986tt}. This approach has been successfully applied in $\cN=4$ super Yang-Mills \cite{Beccaria:2017rbe} to compute the anomalous dimension of the operator interpolating between the ordinary non-supersymmetric Wilson loop and the $1/2$ BPS Wilson-Maldacena loop. 

We begin by considering the Nambu-Goto action \eqref{eq:Nambu-Goto1/2} with Neumann boundary conditions $S_N$ plus a boundary term $\delta S$ 
\begin{equation}
\label{eq:Skappa}
    S(\kappa)_D = S_N + \delta S\,,\qquad \delta S = \kappa\, T \int d\tau Z_I \mathcal{M}^I{}_J \bar Z^J\,.
\end{equation}
The parameter $\kappa$ interpolates between Neumann ($\kappa=0$) and Dirichlet ($\kappa=\infty$)  boundary conditions. 

The boundary term $\delta S$ can be seen as a special case of an open-string tachyon depending on the $\mathbb{CP}^3$ coordinates, for which the $\beta$-function is known. In terms of $\cT = \kappa Z_I \mathcal{M}^I{}_J \bar Z^J$, we have 
\begin{equation}\label{eq:betaT}
    \beta_{\cT} = \mu\frac{\partial \cT}{\partial \mu} = -\cT -\frac{1}{\sqrt{\lambda}}\nabla^2 \cT + \ldots\,,
\end{equation}
where $\nabla^2$ is the Laplace-Beltrami operator on $\mathbb{CP}^3$. 
Therefore, the calculation of the anomalous dimension of the boundary deformations is mapped to the spectral problem for the Laplacian operator in $\mathbb{CP}^3$, which, in the Fubini-Study metric, reads
\begin{equation}
    \nabla^2 = %2
    (1+|\omega|^2)\left( \frac{\partial^2}{\partial \omega_i \partial\bar\omega^i } + \omega_i\frac{\partial}{\partial \omega_i}\bar\omega^j\frac{\partial}{\partial \bar\omega^j} \right)\,.         
\end{equation}
The eigenfunctions of the Laplacian form irreducible representation of $SU(4)$ with Dynkin labels $[p,0,p]$ and eigenvalues $\lambda_p = p(p+3)$. Our composite fields correspond to the $p=1$ case, for which $\lambda_1 = 4$, in agreement with \eqref{eq:ZZ}. For instance, such eigenfunctions could be written as $\cT_I{}^J=\frac{Z_I \bar Z^J}{|Z|^2} -\frac14\delta_J^I$ and, choosing $I=J=1$, we obtain in the Fubini-Study patch $Z_I=(1,w_i)$ ($i=1,2,3$)
\begin{equation}
   -\nabla^2 \cT = -\nabla^2 \left( \frac{1}{1+\omega_i\bar\omega^i} -\frac{1}{4} \right) = 4 \left( \frac{1}{1+\omega_i\bar\omega^i} -\frac{1}{4} \right)
\end{equation}
Going back to \eqref{eq:betaT}, this means that the $\beta$-function for $\kappa$ is
\begin{equation}
    \beta_{\kappa} = \kappa \left(-1 + \frac{4}{\sqrt{\lambda}}+ \ldots\right)\,,
\end{equation}
and, since $ \gamma = \partial_\kappa \beta_\kappa$, we find $\Delta - 1 = -1 + \frac{4}{\sqrt{\lambda}}$, such that
\begin{equation}
    \Delta_{[Z \bar Z]^T} = \frac{4}{\sqrt{\lambda}} + {\cal O}\left(\frac{1}{\lambda}\right)\,.
\end{equation}

More generally, we can design particular boundary deformations to engineer specific RG flows connecting the UV fixed point $W^-$ with notable IR fixed points, such as $W_{1/6}$, $W_{1/2}^+$, and $W^+$. We consider a general quadratic deformation with $\mathcal{M}^I{}_J$ a hermitian matrix. Varying the action \eqref{eq:Skappa} leads to mixed boundary conditions
\begin{equation}
    \partial_n Z_I + \kappa\, \mathcal{M}^J{}_I Z_J = 0 \qquad \text{on } \partial \text{AdS}_2\,,
\end{equation}
where $\partial_n$ denotes the normal derivative at the boundary. Since we established such deformations to be relevant in the UV, then we have $\kappa=0$ in the UV and $\kappa\to\infty$ in the IR. Then, directions corresponding to non-vanishing eigenvalues of $\mathcal{M}^I{}_J$ are driven to Dirichlet boundary conditions in the IR, while those associated with vanishing eigenvalues remain Neumann. In this way, the pattern of boundary conditions, and hence the endpoint of the RG flow, is determined by the structure of $\mathcal{M}^I{}_J$, while the subgroup of $SU(4)$ preserved by the deformation is given by the stabilizer of $\mathcal{M}^I{}_J$.

Notable deformations are obtained by choosing $\mathcal{M}^I{}_J$ to preserve specific subgroups of $SU(4)$. An $SU(4)$ invariant deformation corresponds to
\begin{equation}
    \mathcal{M}^I{}_J = \delta^I_J \,,
\end{equation}
which treats all directions equally and drives all $\mathbb{CP}^3$ coordinates to Dirichlet boundary conditions. The corresponding IR fixed point is $W^+$. 

An $SU(3)$ invariant deformation, on the other hand, can be constructed by selecting a unit vector $n^I$, such as $n^I=(1,0,0,0)$, and defining the projector onto the orthogonal subspace
\begin{equation}
    \mathcal{M}^I{}_J = \delta^I_J - n^I \bar n_J \,,
\end{equation}
which preserves the $SU(3)$ subgroup stabilizing $n^I$. Its IR fixed point is naturally identified as the 1/2 BPS Wilson loop. In fact, for $\kappa\to\infty$ all directions orthogonal to $n^I$ are driven to Dirichlet boundary conditions, while the component along $n^I$ remains Neumann. However, taking into account the projective nature of the homogeneous coordinates, this residual Neumann direction is unphysical, and the endpoint of the string is effectively localized at a point in $\mathbb{CP}^3$, as required for the $1/2$ BPS configuration. This construction captures only the bosonic sector of the deformation. The $1/2$ BPS Wilson loop is not defined solely by a scalar coupling, but it also involves couplings to fermionic degrees of freedom. Consequently, a complete description of the flow to the $W_{1/2}^{+}$ fixed point requires embedding the bosonic boundary term above into a larger deformation admitting a fermionic completion. From the worldsheet perspective, this presumably corresponds to supplementing the bosonic boundary interaction with additional fermionic couplings, which are necessary both to reproduce the full operator structure on the field theory side and to account for the supersymmetry preserved at the endpoint of the flow.

Finally, an $SU(2)\times SU(2)$ invariant deformation is obtained by decomposing $Z^I=(z_i,w_{\hat\imath})$ and choosing
\begin{equation}
    \mathcal{M}^I{}_J = \mathrm{diag}(0,0,1,1)\,,
\end{equation}
which preserves $SU(2)\times SU(2)$ and drives the $w_{\hat\imath}$ directions to Dirichlet boundary conditions while leaving the $z_i$ directions Neumann, in agreement with the structure of the $1/6$ BPS configuration.
\section{The other ordinary Wilson loop, \texorpdfstring{$W^+$}{W+}}
\label{sec:Wplus}
\vskip 7pt

In the previous section we have argued that the dual of the $W^-$ operator is given by Neumann boundary conditions on the $\mathbb{CP}^3$ coordinates, with the smearing performed by integrating over the zero-modes. We are now interested in the dual of $W^+$. In this case, we expect the dual fundamental string description to exhibit the same $SU(4)$ symmetry, but a spectrum of irrelevant deformations, because it must correspond to an IR fixed point of the RG flow. This may be naturally achieved by imposing Dirichlet boundary conditions on $\mathbb{CP}^3$, as hinted at in the previous section. The deformation of $W^-$ leading to $W^{+}$ is distinguished by the fact that it preserves the full $SU(4)$ symmetry, via a boundary term that treats all $\mathbb{CP}^3$ coordinates on an equal footing, without introducing a preferred vector $n^I$ or selecting a specific point in the internal space. Unlike the deformation towards the $1/2$ BPS loop, the endpoint is therefore not associated with a localized string configuration that breaks $SU(4)$ to $SU(3)$. Rather, the full symmetry suggests the presence of a degeneracy parametrizing the position of the string endpoint on $\mathbb{CP}^3$. The natural holographic interpretation is in terms of a string with Dirichlet boundary conditions in the internal directions, supplemented by a smearing over $\mathbb{CP}^3$ that restores the full $SU(4)$ invariance.

The homogeneous $Z_I$ complex coordinates can still be parameterized in an $SU(4)$ covariant manner as
\begin{equation}
    Z_I = \sqrt{1-\zeta^2}\, n_I + \zeta_I \,, \qquad n_I \bar \zeta^I = 0 \,, \qquad n_I \bar n^I = 1\,.
\end{equation}
In the Dirichlet case $n_I$ is not a dynamical variable and is not integrated over in the path integral, in contrast with the Neumann case in \eqref{eq:partition}. It merely specifies the boundary condition, while the fluctuations $\zeta^I$ account for the physical degrees of freedom. Their boundary two-point function is, at a given fixed $n_I$ and up to normalizations, 
\begin{equation}
\label{eq:zeta_prop_dirichlet}
    \left\langle \zeta_I(\tau_1) \bar \zeta^J(\tau_2) \right\rangle_n = \frac{P_I{}^J}{(\tau_1-\tau_2)^2}\,, \qquad P_I{}^J(n) = \delta_I^J - n_I\bar n^J\,.
\end{equation}
In order to restore $SU(4)$ invariance, we eventually average the correlation function over $\mathbb{CP}^3$, thus obtaining
\begin{equation}
    \left\langle \zeta_I(\tau_1) \bar \zeta^J(\tau_2) \right\rangle = \frac{3}{4} \frac{\delta_I^J}{(\tau_1-\tau_2)^2}\,. 
\end{equation}

Bilinear fields could be studied in principle from the calculation of a four-point function
$\langle \zeta_i\bar{\zeta}^j \zeta_k \bar{\zeta}^l \rangle$.
There are, however, two technical obstacles. Unlike the elementary Dirichlet fluctuations of the $1/2$ BPS string, the fields $\zeta_I$ entering these bilinears are not protected, so the corresponding composite operators may acquire anomalous dimensions and mix with other singlets. Moreover, unlike the $W^-$ case with Neumann boundary conditions, the one-loop anomalous dimensions cannot be reduced to the beta function of a simple boundary potential, or equivalently to a simple exchange-diagram calculation. For the Dirichlet construction relevant to $W^+$, one would need the full interacting four-point function of the Dirichlet fluctuations, including contact terms, and the full one-loop anomalous dimension, including possible fermionic contributions. A direct computation would therefore require the quantization of the full Green-Schwarz superstring in the relevant background and the evaluation of the corresponding loop corrections to defect correlators. This is considerably more involved than the protected $1/2$ BPS case and lies beyond the scope of the present work.

\section{Discussion}
\label{sec:conclusions}
\vskip 7pt

In this work we extend the weak-coupling analyses of defect RG flows in ABJM theory initiated in \cite{Castiglioni:2022yes, Castiglioni:2023uus} to the strong coupling regime, using holography. This requires studying string fluctuations in AdS$_4 \times \mathbb{CP}^3$ around a classical AdS$_2$ worldsheet, identifying the combinations of such fluctuations which are dual to the CFT operators triggering the flows, and computing their anomalous dimensions. We find that the qualitative structure of the flows found at weak coupling in \cite{Castiglioni:2022yes, Castiglioni:2023uus} persists also at strong coupling. In particular, we find that
\begin{itemize}
    \item The $1/2$ BPS operator $W_{1/2}^+$ continues to be an IR-stable fixed point also at strong coupling, as it is at weak coupling. The string mode responsible for triggering the flow is the $SU(3)$ invariant bilinear $\omega_i \bar\omega^i$ (with $i=2,3,4$) of massless scalar fluctuations on $\mathbb{CP}^3$ obeying Dirichlet boundary conditions, as it is appropriate for a string solution which is localized at a point of the internal space. The scaling dimension of this irrelevant operator turns out to be
    \begin{equation}
    \Delta_{\omega_i\bar\omega^i} = 2-\frac{3}{\sqrt{\lambda}} +\mathcal{O}\left(\frac{1}{\lambda}\right)\, ,
    \end{equation}
    with the negative sign of the first-order correction suggesting a smooth interpolation with the corresponding weak-coupling result, which is $\Delta=1+$positive correction.
    \item The $1/6$ BPS operator $W_{1/6}$ admits both relevant and irrelevant deformations, matching the saddle-point structure of the dual CFT at weak coupling. In this case, the string modes must preserve an $SU(2)\times SU(2)$ internal symmetry, with the scalar fluctuations splitting in two distinct sets, one obeying Neumann boundary conditions, $z_i$ with $i=1,2$, and one obeying Dirichlet boundary conditions, $w_{\hat\imath}$ with $\hat\imath=3,4$. The Neumann fluctuations give rise to relevant operators, $z_i\bar z^i$ and $[z_i \bar z^j]^T$, with scaling dimensions
    \begin{equation}
    \Delta_{z_i\bar z^i} = \mathcal{O}\left(\frac{1}{\lambda}\right)\, , \qquad \Delta_{[z_i\bar z^j]^T} = \frac{4}{2\sqrt{\lambda}} +  \mathcal{O}\left(\frac{1}{\lambda}\right),
    \end{equation}
    which drive the RG flows away from $W_{1/6}$. On the other hand, the Dirichlet fluctuations combine into irrelevant operators, $w_{\hat \imath}\bar w^{\hat \imath}$ and $[w_{\hat \imath}\bar w^{\hat \jmath}]^ T$, with scaling dimensions
    \begin{equation}
    \Delta_{w_{\hat{\imath}}\bar w^{\hat{\imath}}} = 2-\frac{1}{2\sqrt{\lambda}}+\mathcal{O}\left(\frac{1}{\lambda}\right), \qquad \Delta_{[w_{\hat{\imath}}\bar w^{\hat{\jmath}}]^T} = 2 +\frac{1}{2\sqrt{\lambda}} +  \mathcal{O}\left(\frac{1}{\lambda}\right),
    \end{equation}
    which drive the RG flows toward $W_{1/6}$.
    \item The ordinary, non-supersymmetric loop $W^-$ is a UV fixed point, as in the weak coupling picture. The classical string solution dual to this $SU(4)$ invariant operator is obtained by imposing Neumann boundary conditions on $\mathbb{CP}^3$ and smearing the string over the full internal space. This generalizes what is done for the string dual of the 1/6 BPS loop, which is only smeared over a $\mathbb{CP}^1 \subset \mathbb{CP}^3$. We find that the field composite dual to the CFT operator responsible for the flow is an adjoint combination of the homogeneous coordinates of $\mathbb{CP}^3$, $[Z_I \bar Z^J]^T$ (with $I,J=1,\ldots, 4$). It has scaling dimension
    \begin{equation}
    \Delta_{Z_I \bar Z^I}=0,\qquad 
    \Delta_{[Z_I \bar Z^J]^T} = \frac{4}{\sqrt{\lambda}}+ \mathcal{O}\left(\frac{1}{\lambda}\right),
    \end{equation}
    and corresponds to a relevant deformation away from the UV fixed point. 
\end{itemize}

For convenience, in table \ref{tab:dictionary} we summarize the main identifications between defect-theory deformation operators and worldsheet composite fields used throughout the paper. The entries should be understood as identifications of symmetry channels, and, where mixing is expected, as the lightest operators of definite scaling dimension in the corresponding sector.
\begin{table}[t]
\centering
\renewcommand{\arraystretch}{1.4}
\begin{tabular}{Sc Sc Sc Sc Sc}
\hline
Defect & Channel & Field-theory & Worldsheet & Dimension (to $\mathcal O(\lambda^{-1})$) \\
\hline
$W^+_{1/2}$ 
& $SU(3)$ singlet 
& $S_1,S_2,S_3$ 
& $\omega_i\bar\omega^i$ 
& $2-\dfrac{3}{\sqrt{\lambda}}$ \\
$W^+_{1/2}$ 
& $SU(3)$ adjoint 
& $T^i{}_j$ 
& $[\omega_i\bar\omega^j]^T$ 
& $2-\dfrac{3}{2\sqrt{\lambda}}$ \\
\hline
$W_{1/6}$ 
& $(1,1)$ of $SU(2)\times SU(2)$ 
& $C_i \bar{C}^i$ 
& $z_i\bar z^i$ 
& $\mathcal O(\lambda^{-1})$ \\
$W_{1/6}$ 
& $(3,1)$ of $SU(2)\times SU(2)$ 
& $[C_i \bar{C}^j]^T$ 
& $[z_i\bar z^j]^T$ 
& $\dfrac{2}{\sqrt{\lambda}}$ \\
$W_{1/6}$ 
& $(1,1)$ of $SU(2)\times SU(2)$ 
& $C_{\hat\imath} \bar{C}^{\hat\imath}$   
& $w_{\hat\imath}\bar w^{\hat\imath}$ 
& $2-\dfrac{1}{2\sqrt{\lambda}}$ \\
$W_{1/6}$ 
& $(1,3)$ of $SU(2)\times SU(2)$ 
& $[C_{\hat\imath} \bar{C}^{\hat\jmath}]^T$  
& $[w_{\hat\imath}\bar w^{\hat\jmath}]^T$ 
& $2+\dfrac{1}{2\sqrt{\lambda}}$ \\
\hline
$W^-$ 
& $SU(4)$ adjoint 
& $C_I\bar C^J-\frac14\delta_I^J C_K\bar C^K$ 
& $[Z_I\bar Z^J]^T$ 
& $\dfrac{4}{\sqrt{\lambda}}$ \\
\hline
$W^+$ 
& $SU(4)$ adjoint
& $C_I\bar C^J-\frac14\delta_I^J C_K\bar C^K$ 
& $[\zeta_I\bar\zeta^J]^T$
& not computed explicitly \\
\hline
\end{tabular}
\caption{Summary of the main defect/worldsheet identifications used in the paper. The table records the symmetry channel, the corresponding field-theory deformation, the worldsheet composite, and the strong-coupling dimension when it is explicitly determined. In sectors where operator mixing is expected, the worldsheet operator should be understood as representing the corresponding eigenoperator of definite scaling dimension in that symmetry channel.}
\label{tab:dictionary}
\end{table}

The theory has a second non-supersymmetric Wilson loop, $W^+$ \cite{Castiglioni:2023uus}, for which we propose an holographic dual given by a string with Dirichlet boundary conditions averaged over the whole $\mathbb{CP}^3$, in order to restore the full $SU(4)$ R-symmetry of the dual CFT operator. The difference with $W^-$ is that in that case the zero-modes corresponding to the string orientation in $\mathbb{CP}^3$ are integrated over in the string partition function, whereas for $W^+$ the  orientation in $\mathbb{CP}^3$ is not a dynamical field. As an IR fixed point of the RG flow, we expect $W^+$ to give rise to a spectrum of irrelevant deformations, an analysis that we have not pursued here.

This `doubling' of Wilson loop operators is a peculiar feature of three-dimensional quiver theories: in ABJM theory there are the two non-supersymmetric operators, $W^\pm$, which we have discussed in this work, but there are also two 1/2 BPS operators, $W_{1/2}^\pm$ (of which we have only studied $W_{1/2}^+$).\footnote{Another notable example of this phenomenon is the degeneracy of 1/2 BPS operators in ${\cal N}=4$ Chern-Simons-matter theories, discovered in \cite{Cooke:2015ila} and further studied in \cite{Griguolo:2015swa,Bianchi:2016vvm}.} These pairs of operators only differ in overall signs in the matter couplings, as can be seen for example in table \ref{tab:WLs} for $W^\pm$. It may na\"ively seem that these sign differences are harmless, but they have in fact a profound impact on the properties of these operators under RG flows, as made evident in figure \ref{fig:fluxes}. At weak coupling \cite{Castiglioni:2023uus}, this different behavior can be easily traced back to the sign difference in the definition of the matter couplings percolating through the computation and resulting in crucial sign differences in the beta-functions of the deformation parameters and in the anomalous dimensions of the defect operators. At strong coupling, the connection is less obvious. While we have proposed a dual picture for $W^+$, we have not computed its spectrum of fluctuations. For $W_{1/2}^-$ it is not even clear what the string dual could be. A possibility that could be worth investigating is whether the loops in the unshaded and shaded regions of figure \ref{fig:fluxes} be related by some transformation of the ABJM theory, such as parity perhaps.

Before concluding, let us stress that our analysis focuses on the spectrum of worldsheet fluctuations and the corresponding defect operators at the fixed points. The scaling dimensions obtained in this way allow us to identify relevant and irrelevant deformations and therefore determine the local structure and stability properties of the RG flows. While this provides strong evidence for the pattern of flows connecting the different Wilson loops, we do not derive the full strong-coupling RG trajectories nor the associated beta functions. Constructing such trajectories directly from the holographic description remains an interesting open problem.

Here we have limited our analysis to the bosonic fluctuations around the AdS$_2$ string worldsheet. In a case -- see \eqref{eq:fermioniccontribution} -- we have proposed an explicit value for a fermionic contribution, as needed for consistency, but otherwise  we have always neglected fermionic fluctuations. 
A complete treatment of the unprotected sectors associated with the $W_-$ and $W_{1/2}^+$ configurations would require the inclusion of fermionic worldsheet fluctuations. While the leading identification of the relevant symmetry channels is already captured by the bosonic analysis, fermionic operators may participate in the mixing problem and can modify the anomalous dimensions of unprotected composite operators at one loop and beyond. By contrast, the protected quantities discussed in this work are not expected to be affected qualitatively by the omission of fermions, either because fermionic contributions enter only at higher orders or because supersymmetry fixes the final result. For this reason, the main conclusions regarding the spectrum of relevant and irrelevant deformations remain unchanged, although a complete determination of the unprotected operator spectrum would require an 
analysis of the full Green-Schwarz theory, which we leave for the future. We expect however that the fermions will not change the picture emerged in this work.

Finally, while we have focussed on operators supported along (maximal) circles and transforming in the fundamental representation of the gauge group, it would surely be worthwhile to investigate more general contours, like the latitude of \cite{Cardinali:2012ru,Bianchi:2018bke}, and higher representations, along the lines of the four-dimensional counterpart in \cite{Giombi:2020amn}.

\subsection*{Acknowledgments}
\vskip 3pt
We thank L. Griguolo, G. Korchemsky and A. Tseytlin for useful discussions. SP would like to thank the Isaac Newton Institute for Mathematical Sciences, Cambridge, for hospitality during the programme {\it Quantum Field Theory with Boundaries, Impurities, and Defects}  supported by EPSRC grant no EP/R014604/1, where work on this paper was undertaken. MB is supported by Fondo Nacional de Desarrollo Cient\'ifico y Tecnol\'ogico, through Fondecyt Regular 1220240, Fondecyt Exploraci\'on 13220060 and Fondecyt Exploraci\'on 13250014. MT is supported by the Simons Foundation through award number 1023171-RC and by the Brazilian National Council for Scientific and Technological Development (CNPq) through grant 445944/2024-2. DT is supported in part by FAPESP {\it Tem\'atico} grants 2019/21281-4 and 2024/15298-0 and by the PRD 2026 funds at UniMORE. This work is supported in part by the INFN grant {\it Gauge and String Theory (GAST)}. 

\newpage
\bibliographystyle{unsrt}
\bibliography{refs}
\end{document}